\documentclass[10pt,journal,twoside,final,]{IEEEtran}
\usepackage{cite}
\usepackage{graphicx}
\usepackage{times}
\usepackage{latexsym}

\usepackage[mathscr]{eucal}
\usepackage{array}
\usepackage{fancyhdr}
\usepackage{amsmath,amssymb}
\usepackage{bm} 
\usepackage{subfigure}
\usepackage{multirow}

\ifCLASSINFOpdf
\else
\fi

\newcommand{\diag}{{\rm{diag}}}
\usepackage{xcolor}

\makeatother

\usepackage{comment}
\usepackage{amsthm}
\usepackage{amssymb}
\usepackage{graphicx}
\usepackage{subfigure}
\usepackage{tabularx}
\usepackage{color,cite}
\usepackage{booktabs}
\usepackage{url}
\usepackage{bm}
\usepackage{color}
\definecolor{c}{rgb}{1,0,0} % red
\definecolor{b}{rgb}{0,0,1} % red

\usepackage{algpseudocode}
\usepackage{algorithm}

\usepackage{makecell}
\usepackage{amsfonts}

\usepackage{graphicx}
\usepackage{float}
\usepackage[caption=false,font=footnotesize]{subfig}

\newtheorem{remark}{Remark}
\newtheorem{theorem}{Theorem}
\newtheorem{lemma}[theorem]{Lemma}

\usepackage{algorithm}

\usepackage{algpseudocode}
\allowdisplaybreaks

\begin{document}
	
	\title{Fluid Antenna System-assisted Physical Layer Secret Key Generation}
	
	\author{
		Zhiyu~Huang,
		Guyue~Li,~\IEEEmembership{Member,~IEEE}, Hao~Xu,~\IEEEmembership{Senior Member,~IEEE}, and  Derrick Wing Kwan Ng,~\IEEEmembership{Fellow,~IEEE}
	%and Aiqun~Hu,~\IEEEmembership{Senior Member,~IEEE}

		\thanks{Zhiyu Huang and Guyue Li are with the School of Cyber Science and Engineering, Southeast University, Nanjing 210096, China. Guyue Li is also with Purple Mountain Laboratories, Nanjing 211111, China, and also with the Jiangsu Provincial Key Laboratory of Computer Network Technology, Nanjing 210096, China (e-mail: rain-huang-seu@seu.edu.cn; guyuelee@seu.edu.cn). \textit{(Corresponding author: Guyue Li.)}}
        \thanks{Hao Xu is with the National Mobile Communications Research Laboratory, Southeast University, Nanjing 210096, China, (e-mail: haoxu@seu.edu.cn). }
        \thanks{Derrick Wing Kwan Ng is with the School of Electrical Engineering and Telecommunications, University of New South Wales, Sydney, NSW 2052, Australia (e-mail: w.k.ng@unsw.edu.au).}
		
	%	\thanks{Aiqun Hu is with the National Mobile Communications Research Laboratory, Southeast University, Nanjing, 210096, China, also with Purple Mountain Laboratories, Nanjing 211111, China, and also with the Jiangsu Provincial Key Laboratory of Computer Network Technology, Nanjing 210096, China (e-mail: aqhu@seu.edu.cn).} 
		}

	\maketitle
	
	\begin{abstract}
     This paper investigates {physical-layer} key generation (PLKG) in multi-antenna base station (BS) systems, {by leveraging} a fluid antenna system (FAS) to dynamically {customize} radio environments. Without {requiring} additional nodes or extensive radio frequency chains, {the} FAS {effectively} enables adaptive antenna port {selection by exploiting} channel spatial correlation {to enhance} {the} key generation rate (KGR) at legitimate nodes. 
     To {comprehensively evaluate the efficiency} of {the} FAS {in} PLKG, we propose {an} FAS-assisted PLKG model that integrates transmit beamforming and sparse port selection under independent and identically distributed (i.i.d.) and spatially correlated channel models, respectively. {Specifically}, the PLKG {utilizes} reciprocal channel probing to derive a closed-form KGR expression based on {the} mutual information between legitimate channel estimates, explicitly accounting for Eve's channel observation {under} spatially correlated channel {scenarios}. 
     Nonconvex optimization problems for {these} scenarios are formulated to maximize {the} KGR subject to transmit power constraints and sparse port activation. 
     {We propose an iterative algorithm by capitalizing {on} successive convex approximation (SCA) and Cauchy-Schwarz inequality to obtain {a} locally optimal solution.} {A reweighted $\ell_1$-norm-based algorithm is applied to {advocate} {for} the sparse port activation of FAS-assisted PLKG.} {To approximate the optimal activated ports obtained by exhaustive search},
     {a low-complexity sliding window-based port selection is proposed to substitute reweighted $\ell_1$-norm method based on {Rayleigh-quotient analysis.}}
     Simulation results demonstrate that the FAS-PLKG scheme significantly {outperforms the} FA-PLKG scheme in both independent and spatially correlated {environments}. Furthermore, the sliding window-based port selection method {introduced} in this paper has been shown to {yield superior KGR, compared to} the reweighted $\ell_1$-norm method. {It is shown that the FAS achieves higher KGR with fewer RF chains through dynamic sparse port selection, which effectively reduces the resource overhead. Also, the sliding window approach proposed in this paper {closely} approximates the globally optimal port selection compared to the reweighted $\ell_1$-norm method, {rendering it suitable} for practical deployments.}
	\end{abstract}
	
	\begin{IEEEkeywords}
		Physical layer security, secret key generation, fluid antenna system, spatially correlated channels.
	\end{IEEEkeywords}
	
	\section{Introduction}
        The exponential growth of wireless communication systems has {led to} an unprecedented demand for secure frameworks to protect sensitive data transmitted over shared wireless channels \cite{6GKAF2024}. In dynamic, decentralized {ad-hoc} networks such as { the Internet-of-Things (IoT)}, traditional cryptographic methods relying on pre-distributed keys {encounter} critical limitations, including constrained computational resources, frequent {network} topology {variations}, and impractical centralized key management \cite{PLS2023}. {To this end}, physical-layer key generation (PLKG) addresses these issues by leveraging inherent {and unique} wireless channel {characteristics}, {such as} randomness, reciprocity, and spatial decorrelation, to establish shared secret keys without relying on {pre-existing} infrastructure \cite{Maurer1993}. 
        By extracting entropy from channel state information (CSI), such as phase shifts \cite{EKB21}, received signal strength (RSS) \cite{ZZHX22} and multipath delays \cite{DTAE23}, PLKG ensures that legitimate transceivers {generate} correlated keys {owing} to their {common} propagation environment, while eavesdroppers at different locations observe decorrelated channel responses \cite{LCHW17}. {In fact,} this spatial decorrelation {establishes} the {foundation} of information-theoretic security, preventing adversaries from reliably estimating keys, even with unbounded computational resources. 
        Unlike traditional methods, PLKG eliminates the need for secure key distribution channels and complex {cryptographic} operations, {rendering} it {particularly suitable} for IoT {scenarios} \cite{GCXM21}. {For instance, through} channel reciprocity, legitimate parties derive identical keys through bidirectional {channel} probing, while spatial diversity guarantees {the uniqueness and unpredictability of these keys}. {Furthermore,} PLKG has been validated in practical systems, demonstrating its feasibility to secure {wireless} communications \cite{JWWAT19}. 
        
        {Despite these advantages,} PLKG remains challenging due to inherent limitations {associated with} wireless propagation and adversarial environments. {Specifically, the efficiency of} PLKG {heavily} relies on {channel} reciprocity and dynamic fading characteristics to extract the secure randomness \cite{2021Sum}. 
        However, practical implementations face significant obstacles in scenarios {involving} shadowed environments or static channels, where {insufficient} channel entropy {undermines} reliable key generation \cite{BSPD19}. 
        {Additionally,} directional beamforming introduces spatial sparsity that {further} reduces the available randomness \cite{WFLY25}. 
        Moreover, conventional approaches struggle to maintain security in {scenarios} with correlated eavesdropping channels or devices with limited resources, where pre-shared keys are impractical \cite{LWJY23}. 
        Recent {advancements have mitigated} these challenges through innovative architectural and signal processing techniques. 
        {Particularly}, in traditional multiple {fixed}-antenna (FA) communication systems, {beam-domain} channel modeling leverages spatial decorrelation to isolate legitimate users from eavesdroppers. This approach {significantly} reduces pilot overhead and enhances security by exploiting sparse channel representations in {the} beamspace \cite{LMZW25}. 
        Furthermore, {combining} hybrid beamforming with channel estimates {enables systems to dynamically adapt}, {compensating for} {any potential} reciprocity mismatches and improving key-generation {robustness} \cite{LXXJH21}. 
        However, FA-based approaches {{typically} require numerous} radio frequency chains, thereby increasing deployment complexity and communication {overhead}.

        {To overcome these issues}, {a} fluid antenna system (FAS) {have been introduced to} dynamically reconfigure antenna positions {in practical scenarios}, {{thereby} effectively exploiting spatial diversity as a {versatile} alternative to traditional fixed antenna arrays \cite{YXWJWY25}.} {Specifically, by} adaptively selecting optimal ports via software control, {the} FAS enhances channel gains and {achieves improved} performance with {a reduced number of} radio frequency chains \cite{NWXT24}. This agility {effectively lowers} hardware complexity and energy consumption compared to fixed arrays, {positioning} {the} FAS {as an ideal solution} for {enabling} next-generation wireless networks.
        {Due to its tremendous potential,} {the notion of} FAS has been {applied to} physical-layer secure communication through adaptive antenna positioning and jamming strategies \cite{VUMG24}. Recent studies {have demonstrated} the {excellent} ability of {the} FAS to maximize {the} secrecy {rate} by optimizing antenna positions to strengthen legitimate channels while disrupting eavesdroppers \cite{TXWTZ23}. For instance, coding-enhanced cooperative jamming in {the} FAS {has been shown to} outperform conventional Gaussian noise jamming by {exploiting} spatial diversity to {effectively harness} and manipulate interference \cite{XWXC24}. Additionally, FAS-aided systems achieve robust secrecy performance under correlated fading channels, highlighting their potential for {securing wireless communication} \cite{RWJK24}.
    
        {On the other hand,} {in wireless IoT networks,} PLKG is tailored for {resource-constrained} devices. {However,} traditional PLKG designs often demand numerous RF chains, overburdening these devices' hardware resources\cite{MA2015,WWAJZ19,JWWAT19}. {Motivated} by the above studies on FAS, the integration of FAS and PLKG potentially addresses a fundamental limitation {in} {lightweight} IoT {deployments}. {Indeed,} {employing the} FAS can address this issue by dynamically reconfiguring activated ports with fewer RF chains\cite{NWXTC24,CCWS25,MCWZ25}. {In particular,} with the requirement for few radio frequency chains, the dynamic spatial reconfiguration of FAS is expected to increase the entropy of key extraction at the legitimate node. {In addition, }the larger number of {potential} FAS preset ports also {introduces} {additional} randomness to the PLKG compared to traditional fixed antennas\cite{WSTZ21}. 
        {Nonetheless, existing multi-antenna PLKG strategies, e.g. \cite{WFLY25,LWJY23}, require complex processing in the beam domain to improve the KGR. In addition, current security strategies for FAS, e.g. \cite{TXWTZ23,XWXC24,RWJK24}, {primarily} consider Wyner-based degenerate eavesdropping channel assumptions and {neglect} secure communication {scenarios where} the eavesdropper's channel {may} outperform that of the legitimate receiver.}
        {Thus,} the impact of FAS on PLKG {still remains unexplored}. 
        {As a matter of fact, addressing this gap requires developing sophisticated modeling techniques and deriving accurate expression for the key generation rate applicable to various scenarios.}
        Moreover, it is imperative to surmount the obstacles inherent in non-convex optimization, particularly {related to} sparse port selection and beamforming {design}. Optimal port selection is a combinatorial problem with a $\ell_0$-norm sparsity, which is computationally intractable for the problem. These challenges necessitate innovative solutions {that bridge} information theory and optimization {tailored to efficient} secret key generation.
    
        To investigate FAS-assisted PLKG in a {multiple-input-single output (MISO) system, this paper considers both i.i.d. and spatially correlated channel scenarios}. {Our main contributions {are} summarized as follows.}

    \begin{itemize}
        \item {We propose two FAS-assisted PLKG models based on} {an} independent {scenario,} where Eve's estimate is independent of the legitimate {users'} estimates, and {a} correlated {scenario} {that incorporates} spatial correlation between eavesdropping and legitimate channels. Closed-form KGR expressions are derived for both scenarios {by exploiting} mutual information and joint entropy analysis, {considering {both} the beamforming and FAS correlation channel model at {the} BS. Additionally, our analysis shows that the KGR gain increases with the largest eigenvalue of {the} spatially correlated matrix at {the} BS.} 
        
        \item To maximize a tractable surrogate function, the nonconvex {terms} {in the formulated problem} are addressed via successive convex approximation (SCA) combined with {the} Cauchy-Schwarz inequality relaxation. To {promote} { sparsity in the solution to port activation problem, we propose} a reweighted $\ell_1$-norm method. {Moreover, to approximate the optimal activated ports obtained by exhaustive search,} a sliding window-based port selection is proposed {to improve port activation guided by} Rayleigh-quotient theory, {significantly reducing computational complexity.}
        
        \item  {Simulation results demonstrate that the {proposed} FAS-PLKG scheme significantly outperforms {the} FA-PLKG scheme in both independent and spatially correlated scenarios. {With the same order of complexity, the sliding window-based port selection method {consistently achieves} superior KGR {performance}, compared to the reweighted $\ell_1$-norm method. In addition, the sliding window approach {closely} approximates the optimal port selection obtained by traversing all ports with exceedingly high computational complexity.}}
    \end{itemize}
	\emph{Notations:} 
	In this paper, matrices and vectors are denoted by boldface upper-case and boldface lower-case, {respectively}. 
	$\mathbb{C}^{A\times B}$ denotes the space of complex matrices of size $A\times B$. 
    {$\mathcal{O}(\cdot)$ denotes the complexity.}
	%$\Re(\cdot)$ and $\Im(\cdot)$ stand for the real and imaginary parts of a complex number. The imaginary unit of a complex number is denoted by $j=\sqrt{-1}$.
	$(\cdot)^*$, $(\cdot)^{\top}$, and $(\cdot)^{\sf H}$ denote the conjugate, transpose, and conjugate transpose, respectively. $\diag(\boldsymbol{x})$
	is a matrix whose main diagonal elements are the entries of $\boldsymbol{x}$.
    {$\frac{d f(x)}{d x}$ denotes that the function $f(x)$ takes the derivative of $x$}.
	%$\text{vec}(\boldsymbol{X})$ denotes the vectorization of the matrix $\boldsymbol{X}$. 
	% $\tr(\cdot)$ and $\text{rank}(\cdot)$ represent the trace and rank of a matrix. 
	% $\triangleq$ means “defined as”. 
	%The Kronecker product, Hadamard product, and Khatri-Rao product are represented by $\otimes$, $\circ$, and $\odot$, respectively. $\mathcal{I}(X;Y)$ and $\mathcal{H}(X,Y)$ are the mutual information and joint entropy of random variables $X$ and $Y$, respectively. 
	%$\mathcal{H}(X,Y|Z)$ is the conditional entropy of $X$ and $Y$ given $Z$.
	$\operatorname{det}(\cdot)$ is the matrix determinant. $||\boldsymbol{x}||_{0}$ , $||\boldsymbol{x}||_{1}$, and $||\boldsymbol{x}||_{2}$ denote the $\ell_{0}$, $\ell_{1}$, and $\ell_{2}$ norms of vector $\boldsymbol{x}$, {respectively}. 
	$\mathbb{E}\{\cdot\}$ represents statistical expectation. 
	$\lambda_{\mathrm{max}}(\boldsymbol{X})$ is the maximum eigenvalue of matrix $\boldsymbol{X}$.
        $\boldsymbol{b}\sim \mathcal{CN}\left(\boldsymbol{0}, \boldsymbol{\Sigma}\right) $ denotes that $\boldsymbol{b}$ is a circularly
	symmetric complex Gaussian (CSCG) vector with zero mean and
	covariance matrix $\boldsymbol{\Sigma}$.
        {${\rm Cov}[a,b]$ denotes {the} covariance of {random variables} $a$ and $b$.
        $\mathcal{I}\left(X\mid Y\right)$ is the mutual information between $X$ and $Y$.
        $J_0(\cdot)$ is the zero-order Bessel function of the first kind.}
        {$\mathrm{Re}\{\cdot\}$ represents the real part of a complex number.}

        %%$\mathcal{O}(\cdot)$ is the big-O notation.
	%%$\boldsymbol{X} \succeq \boldsymbol{0}$ means $\boldsymbol{X}$ is a positive semidefinite matrix.
	%%$\nabla f(\cdot)$ and 
	%%$\partial f/\partial x$ are the gradient operators of function $f$. $\boldsymbol{I}$ denotes the identity matrix. 

\section{System Model}\label{sec:system_model}
	\begin{figure}[ht]
		\centering
		\includegraphics[width=2.7in]{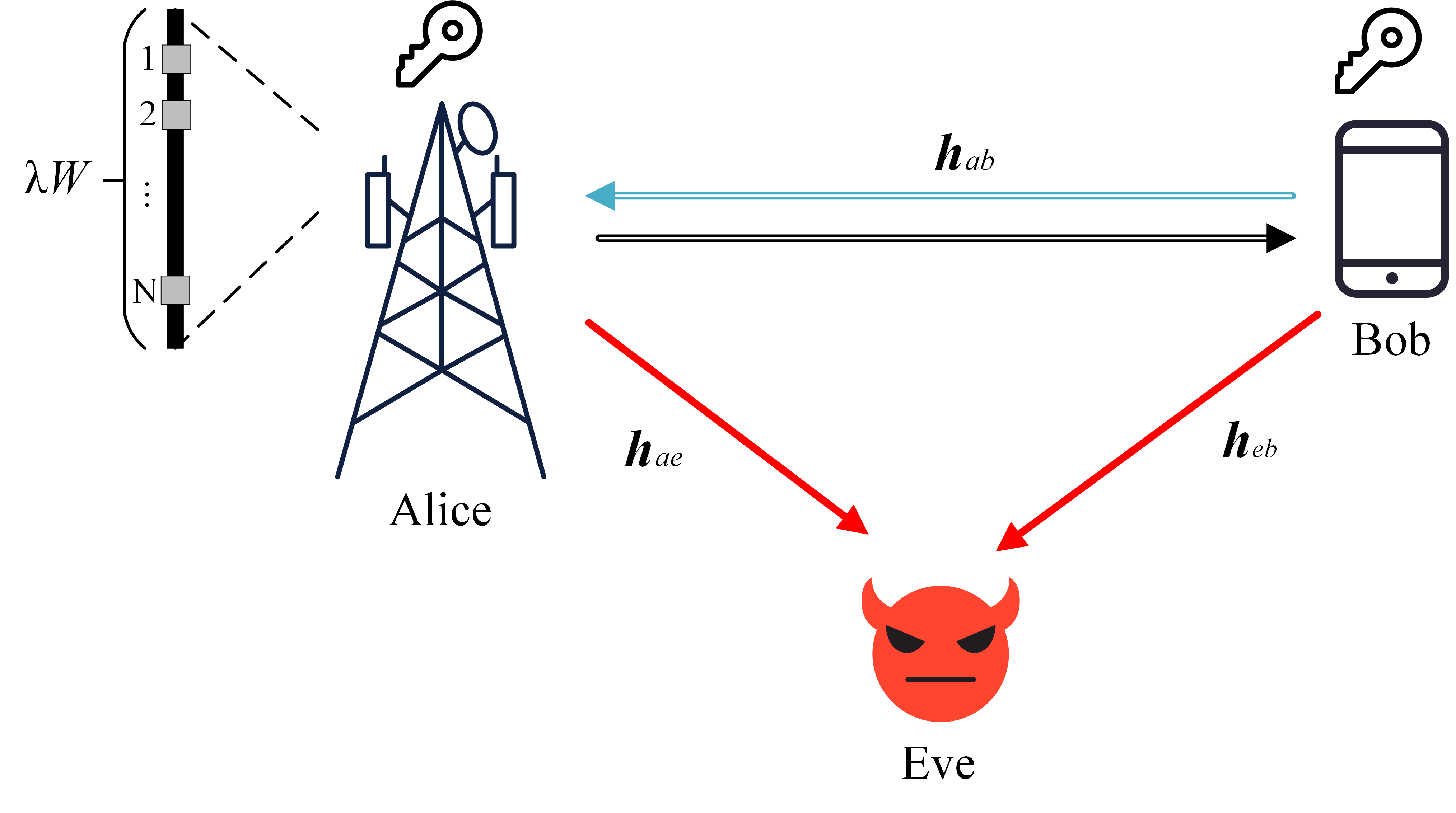}
		\caption{The model of FAS-assisted secret key generation in MISO systems.}
		\label{system model}
	\end{figure}
	As shown in Fig.~\ref{system model}, we {investigate} {a MISO system implementing} {an} FAS-assisted PLKG {scheme}. Assuming a time-division duplexing (TDD) protocol is adopted, a multi-antenna base station (BS), Alice, and a single fixed-position antenna user, Bob, aim to generate symmetric keys by exploiting the reciprocity of the wireless channel with {the assistance of an} FAS. Meanwhile, a single fixed-position antenna eavesdropper, Eve, {attempts} to {intercept and decode} the secret key information {contained within} the received signals{\footnote{Note that the considered framework can be generalized to multiple single-antenna eavesdroppers by optimizing against the worst case scenario (i.e., the eavesdropper with the strongest channel conditions)}}. 
	%Specifically, assuming a time-division duplexing (TDD) mode, Alice and Bob sound channel alternatively in coherence time to obtain the reciprocal channel estimate.
	\subsection{Channel Model}
    In FAS-assisted PLKG, we assume that Alice is equipped with a {linear} fluid antenna surface {with} $ N $ RF chains, which is regarded as a common case of FAS \cite{WSTZ21}. $M$ predetermined ports share $ N $ RF chains and are uniformly distributed along a linear space of length $\lambda W$, where $\lambda$ is the {signal} carrier wavelength and $W$ is the normalized size of FAS. 
    The $N$ RF chains can {adaptively alter} their activated ports among the $M$ available pre-set ports such that the PLKG performance can be enhanced\footnote{The rationale of FAS resembles traditional antenna selection. From the signal processing viewpoint, they are indeed similar. However, it is worth noting that FAS is designed to switch the antenna's position very finely within a given space, typically handling an enormous number of correlated signals for selection\cite{WSTZ20,NWXW24}. }. Compared with the traditional MISO {systems}, the port switching capability of FAS {offers} more spatial DoF to the PLKG BS\cite{WSTZ21}.
	Since the spatial correlation affects the secret key rate, we consider the general spatial correlation channel model at Alice.
	The direct channels of Alice-to-Bob, Eve-to-Bob, and Alice-to-Eve are denoted by $\boldsymbol{h}_{\rm{ab}}  \in \mathbb{C}^{M \times 1}$, 
	${h}_{\rm{eb}}\in \mathbb{C}^{1 \times 1}$, and $\boldsymbol{h}_{\rm{ae}}  \in \mathbb{C}^{M \times 1}$, respectively. 
	To account for the spatial correlation\footnote{
    %However, for traditional antenna selection systems, multiple antennas are deployed at fixed locations and the antenna with the strongest signal is selected. By contrast, in FAS, there is only one antenna or fewer antennas whose position (referred to as “port”) is flexible within a predefined space. 
    %FAS selects the best port or port combination for the strongest reception in the same way as choosing the best antennas in a multiple antenna system\cite{NWXW24}. 
    %The geometric parameters of radiating elements—including their positions, orientations, or lengths—can be reconfigured either discretely or continuously in fluid antenna systems. 
    %In the current stage, the emphasis of FAS primarily centers around position reconfiguration\cite{YXWJWY25} and port selection\cite{CWTK22}. 
     To accurately capture the intricate spatial dependencies between antenna ports, our investigation employs a fully correlated channel model\cite{KKAM23}, which effectively models the mutual coupling and propagation dynamics inherent in reconfigurable antenna arrays. This approach ensures the spatial correlation effects, which are critical for optimizing PLKG.}, the channel matrices are described by employing the FAS correlation channel model \cite{NWXTC24} as 
	$\boldsymbol{h}_{{\rm ab}}= \beta_{{\rm ab}}^{\frac{1}{2}}\boldsymbol{J}_{\rm A}^{\frac{1}{2}} \tilde{ \boldsymbol{h}}_{{\rm ab}}$, $\boldsymbol{h}_{{\rm ae}}= \beta_{{\rm ae}}^{\frac{1}{2}}\boldsymbol{J}_{\rm A}^{\frac{1}{2}} \tilde{ \boldsymbol{h}}_{{\rm ae}}$
    respectively, where $\boldsymbol{J}_{\rm A} \in \mathbb{C}^{M \times M}$ is the spatial correlation matrix at Alice\cite{NWXW24,Jakes}. The $(n,m)$-th element of $\boldsymbol{J}_{\rm A}$ can be expressed as
    \begin{align}
        \boldsymbol{J}_{{\rm A}}(n,m)={\rm Cov}[h_n,h_m]=J_0\left(2\pi\frac{|n-m|}{M-1}W\right).
    \end{align}
    In addition, 
	$\tilde{ \boldsymbol{h}}_{{\rm ab}}\in \mathbb{C}^{M \times 1}$ and $\tilde{ \boldsymbol{h}}_{{\rm ae}}\in \mathbb{C}^{M \times 1}$ are random matrices with i.i.d. Gaussian random entries of zero mean and unit variance. 
	$\beta_{\rm ab}$ and $\beta_{{\rm ae}}$ are the path {losses} of the corresponding channels.

	\subsection{PLKG System Based on Transmit Beamforming}
    {
	In this section, we propose a PLKG framework based on the established channel model of FAS. In the considered FAS-assisted PLKG system, Alice and Bob acquire reciprocal channel estimates by {performing} channel probing. The detailed {procedure} is introduced as follows\cite{LCZCH23,HLQH24}.   
	
	In the downlink stage, initiated by Alice's transmission of a predefined pilot sequence \( s_{\rm d} \in \mathbb{C} \) with unit power (\(|s_{\rm d}|^2 = 1 \)), the received signals at Bob and Eve {are} modeled as:
	\begin{align}
		{y}_{\rm l}^{\rm d} &=  \boldsymbol{h}_{\rm al}^{\top}\boldsymbol{w} s_{\rm d} +  {z}_{\rm l}^{\rm d},{\rm l}\in \{{\rm b},{\rm e}\}, \label{eq:RIS_model} 
	\end{align}
    where $\boldsymbol{w}\in\mathbb{C}^{M\times 1}$ denotes the sparse beamforming vector at Alice, {satisfying} $\|\boldsymbol{w}\|_{0} = N$ and $\|\boldsymbol{w}\|_{2}^2 \leq P_\mathrm{A}$, with \(P_\mathrm{A}\) {denoting} the maximum transmit power budget. {Also,} ${z}_{\rm l}^{\rm d} $ is the additive {white} Gaussian noise component, satisfying ${{z}}_{\rm l}^{\rm d} \sim \mathcal{CN}\left(0, \sigma_{\rm l}^2 \right)$ {with $\sigma_{\rm l}^2$ {being} the received antenna noise power.} 
     
    Subsequently, both receivers employ least-squares (LS) estimation techniques \cite{21JiTVT,22Sum-RIS} to derive channel state information\footnote{The LS is a common technique to obtain estimates in practical systems\cite{2020Intelligent}.}:
	\begin{align}
		\hat{ h}_{ \rm b }\triangleq  s_{\rm d}^*y_{\rm b}^{\rm d} 
		&= \boldsymbol{h}_{\rm ab}^{ \top } \boldsymbol{w} + \tilde{z}_{\rm b}^{\rm d},  \label{hb} \\
		\hat{{h}}_{\rm  e }^{\rm d} \triangleq  s_{\rm d}^* {y}_{\rm e}^{\rm d} 
		&= \boldsymbol{h}_{\rm ae}^{\top} \boldsymbol{w} + \tilde{{z}}_{\rm e}^{\rm d},
	\end{align}
	where \(\tilde{z}_{\rm b}^{\rm d } = s_{\rm d}^* z_{\rm b}^{\rm d} \) and \(\tilde{{z}}_{\rm e}^{d}=s_{\rm d}^*{z}_{\rm e}^{\rm d}\) represent the noise terms, respectively.
	
	During {the} uplink stage, Bob transmits {the} corresponding uplink pilot $s_{\rm u} \in\mathbb{C} $ with $|s_{\rm u}|^2=1$, yielding received signals at Alice and Eve: 
	\begin{align}
			\boldsymbol{y}_{\rm l}^{\rm u} = \sqrt{P_\mathrm{B}} \boldsymbol{h}_{\rm l{\rm b}}     s_{\rm u} + \boldsymbol{z}^{\rm u}_{\rm l}, {\rm l}\in \{{\rm a},{\rm e}\},
	\end{align}
    where $P_\mathrm{B}$ denotes Bob's transmit power budget. The noise components maintain \(\boldsymbol{z}^{\rm u}_{\rm l} \sim \mathcal{CN}(0, \sigma_{\rm l}^2\bm{I}_M)\). 
    
    Through LS {estimation}, the channel estimates become:
	\begin{align}
		\hat{\boldsymbol{h}}_{\rm l}^{\rm u} \triangleq s_{\rm u}^*\boldsymbol{y}_{\rm l}^{\rm u} 
		= \sqrt{P_\mathrm{B}} \boldsymbol{h}_{{\rm l}{\rm b}} + \tilde{\boldsymbol{z}}_{{\rm l}}^{\rm u}, {\rm l}\in \{{\rm a},{\rm e}\}, \label{ha1} 
	\end{align} 
	where $\tilde{\boldsymbol{z}}_{\rm l}^{u}=s_{\rm u}^*\boldsymbol{z}_{\rm l}^{\rm u}$. 
	
    {To align Alice's vector estimate} \(\hat{\boldsymbol{h}}_{\rm a}^{\rm u}\) {with} Bob's scalar estimate \(\hat{h}_{\rm b}\), we construct an equivalent channel measurement at Alice {by applying} beamforming:
	\begin{align}
		\hat{ h}_{\rm a}  \triangleq  \boldsymbol{w}^{\sf H}  \hat{\boldsymbol{h}}_{\rm a}^{\rm u}= \sqrt{P_\mathrm{B}} \boldsymbol{w}^{\sf H} \boldsymbol{h}_{\rm ab} + z^{\rm u}_{\rm a}, \label{ha}
	\end{align}
	where the noise is $z^{\rm u}_{\rm a}=\boldsymbol{w}^{\sf H} \tilde{\boldsymbol{z}}_{\rm a}^{u}$. 
	
    {In practice,} the constructed channel parameters \(\hat{h}_{\rm a}\) and \(\hat{h}_{\rm b}\) exhibit strong temporal correlation within channel coherence intervals \cite{2021Sum,li2018high,ZhangReview}. Following standard PLKG procedures including quantization, information reconciliation, and privacy amplification \cite{2021Sum}, these correlated measurements are converted into secret keys. Distinct from conventional approaches \cite{li2018high,ZhangReview}, our contribution {emphasizes} on optimizing the channel probing phase through joint beamforming design and port selection to maximize the {resulting} KGR.
    }

	\section{Problem Formulation}
	In this section, {leveraging} the channel {estimates} acquired in Sec. \ref{sec:system_model}, 
	we formulate an optimization problem {over the beamforming vector} $\boldsymbol{w}$ to improve system performance.
	{To this end,} we first {derive} the KGR given Eve's channel estimates that facilitate its subsequence maximization design {through optimization}\cite{21JiTVT}. 
	Specifically, {given the observation of Eve's antenna}, the KGR is defined as the conditional mutual information of the legitimate parties' channel estimates\cite{li2018high} as follows\footnote{In this paper, we focus on the investigation of PLKG under the {far-field} plane-wave assumption {for} FAS, {considering} two common eavesdropping presence scenarios \cite{22Globecom,HLQH24}.} 
    
	\begin{align}\label{Rsk_ini}
	R_{\rm SK} = \mathcal{I}\left(\hat{ h}_{\rm a} ; \hat{ h}_{\rm b}\mid \hat{ h}_{{\rm e}}^{\rm d},\hat{ h}_{{\rm e}}^{\rm u}\right).
	\end{align}

{Subsequently, we establish two FAS-assisted PLKG models based on an independent scenario, where Eve’s estimate
is independent of the legitimate users’ estimates, and a correlated scenario that incorporates spatial correlation between eavesdropping and legitimate channels. Closed-form KGR expressions are derived for both scenarios by exploiting mutual information and joint entropy analysis, considering both the beamforming and FAS correlation channel model at the BS.}
    
    \subsection{Independent and Identically Distributed Scenario}
	In {the} i.i.d scenario, Eve's channel is independent of {the channels between Alice and Bob}. {Under this assumption}, the KGR is given by \cite{jorswieck2013secret},\cite{wong2009secret}
	\begin{align}
		R^{\rm iid}_{\rm SK}
	  =\mathcal{I}\left(\hat{ h}_{\rm a} ; \hat{ h}_{\rm b}\right)
		= \log_{2} \frac{\mathcal{R}_{{\rm a}{\rm a}} \mathcal{R}_{{\rm b}{\rm b}}}{\operatorname{det}\left(\boldsymbol{R}_{{\rm ab}} \right)} . \label{eq:MI}
	\end{align}
    
	The covariance {matrix for the channel estimates at} Alice and Bob is defined as follows
	\begin{align}
		\boldsymbol{R}_{{\rm ab}} =\left[\begin{array}{ll}
			\mathcal{R}_{\rm aa} & \mathcal{R}_{{\rm ab}}  \\
			\mathcal{R}_{{\rm ba} } & \mathcal{R}_{{\rm bb}} 
		\end{array}\right],
	\end{align}
	where $\mathcal{R}_{\rm xy}=\mathbb{E}\left\{\hat{ h}_{\rm x}\hat{ h}_{\rm y}^{\sf H} \right\}, {\rm x},{\rm y} \in \{ {\rm a},{\rm b}\}$ 
	is the channel covariances of the corresponding channel estimates. 
    
	By incorporating the derived channel estimates into the mutual information expression (\ref{eq:MI}) and assuming uniform noise variance (\(\sigma_{ \rm a }^2 = \sigma_{ \rm b }^2 = \sigma_{ {\rm e} }^2 \triangleq \sigma^2\)) {, as commonly adopted in} \cite{21JiTVT,22Sum-RIS}, we establish a closed-form expression for the KGR through the following lemma.
	\begin{lemma}
	The KGR between Alice and Bob is expressed as follows
%The KGR between Alice and Bob is expressed as follows
		\begin{equation}\label{eq:whole}
	   R^{\rm iid}_{\mathrm{SK}}=\log _{2} \frac{\left(P_{\mathrm{B}} \boldsymbol{w}^{\sf H} \boldsymbol{J}_{\rm A} \boldsymbol{w}\beta_{\rm ba}+\|\boldsymbol{w}\|_2^{2} \sigma^{2}\right)\left(\boldsymbol{w}^{\sf H} \boldsymbol{J}_{\rm A} \boldsymbol{w}\beta_{\rm ba}+\sigma^{2}\right)}{\left(\|\boldsymbol{w}\|_2^{2}+P_{\mathrm{B}}\right) \sigma^{2} \boldsymbol{w}^{\sf H} \boldsymbol{J}_{\rm A} \boldsymbol{w}\beta_{\rm ba}+\|\boldsymbol{w}\|_2^{2} \sigma^{4}} .
		\end{equation}
    \end{lemma}
    
	\begin{IEEEproof}
		{Please refer to} Appendix~\ref{sec:covariance_calculation}. 
	\end{IEEEproof}

	Thus, the beamforming design could be formulated as
    \begin{subequations}\label{problem:P1}
    \begin{align}
		{\rm P1:}\max _{\boldsymbol{w}} \
		&R^{\rm iid}_{\rm SK}  \\
		\ \text { s.t. } 
		&\ \text {C1}{:}\,||\boldsymbol{w}||_{2}^2 \leq P_\mathrm{A} , \\
		&\ \text {C2}{:}\,||\boldsymbol{w}||_{0} \leq N , 
	\end{align}  
    \end{subequations}
    where {constraint} $\text {C1}$ indicates that {the} transmit beamforming {power does not exceed} the maximum transmit power budget $P_{\rm A}$, while $\text {C2}$ represents the constraint {on the number of available} RF chains. 
	
	%\subsection{Problem Transformation}
	It could be {observed} that the {optimization} problem {in} (\ref{problem:P1}) is highly nonconvex due to the high-order {coupling of optimization variables} in the objective function. In addition, the port selection is non-convex combinatorial {in nature} due to the sparse constraint, which makes problem (\ref{problem:P1}) more challenging to handle.

\subsection{Spatially Correlated Eavesdropping Scenario}
In this section, {we assume} that {Eve's downlink} channel estimates {are} correlated with {those of legitimate parties}. Based on \eqref{Rsk_ini}, the KGR can be given by
	\begin{align}
		R^{\rm cc}_{\rm SK}
	  =\mathcal{I}\left(\hat{ h}_{\rm a} ; \hat{ h}_{\rm b}\mid \hat{ h}_{{\rm e}}^{\rm d}\right)
	= \log_{2} \frac{\operatorname{det}\left(\boldsymbol{R}_{{\rm a}{\rm e}} \boldsymbol{R}_{{\rm b}{\rm e}}\right)}{\operatorname{det}\left(\boldsymbol{R}_{{\rm abe}} \right)\mathcal{R}_{\rm ee}} , \label{eq:Rsk_cc_ini}
	\end{align}
where the covariance matrices are defined as
	\begin{align}
		\boldsymbol{R}_{{\rm ue}} &=\left[\begin{array}{ll}
			\mathcal{R}_{\rm uu} & \mathcal{R}_{{\rm ue}}  \\
			\mathcal{R}_{{\rm eu} } & \mathcal{R}_{{\rm ee}} 
		\end{array}\right], {\rm u} \in \{\rm a,b\},\\
    	\boldsymbol{R}_{{\rm abe}} &=\left[\begin{array}{lll}
		\mathcal{R}_{\rm aa} & \mathcal{R}_{{\rm ab}} & \mathcal{R}_{{\rm ae}} \\
		\mathcal{R}_{{\rm ba} } & \mathcal{R}_{{\rm bb}} & \mathcal{R}_{{\rm be}}\\
            \mathcal{R}_{{\rm ea} } & \mathcal{R}_{{\rm eb}} & \mathcal{R}_{{\rm ee}}
		\end{array}\right].
    \end{align}

By {exploiting} {\textbf{Lemma~1}}, the KGR between Alice and Bob, given the estimate of the channel in Eve, can be given by \eqref{eq:whole_cc}, shown at the top of next page, where $H_\mathrm{u}=\beta_{\rm ab}\boldsymbol{w}^{\sf H} \boldsymbol{J}_{\rm A} \boldsymbol{w}$, $H_\mathrm{e}=\beta_{\rm ae}\boldsymbol{w}^{\sf H} \boldsymbol{J}_{\rm A} \boldsymbol{w}$, and $H_\mathrm{ue}=\sqrt{\beta_{\rm ab}\beta_{\rm ae}}\boldsymbol{w}^{\sf H} \boldsymbol{J}_{\rm A} \boldsymbol{w}$.
	%%------------跨栏公式
	\newcounter{mytempeqcnt}
	\begin{figure*}[!t]
		\normalsize
\begin{align}\label{eq:whole_cc}
&R^{\rm cc}_{\mathrm{SK}}=\notag\\
&\log_2\left(\frac{\left(P_\mathrm{B}H_\mathrm{u}H_\mathrm{e}
\!+\!(P_\mathrm{B}H_\mathrm{u}
\!+\!H_\mathrm{e}||\boldsymbol{w}||_2^2)\sigma^2
\!+\!||\boldsymbol{w}||_2^2\sigma^4
\!-\!P_\mathrm{B}|H_\mathrm{ue}|^2\right)\left(H_\mathrm{u}H_\mathrm{e}
\!+\!(H_\mathrm{u}
+H_\mathrm{e})\sigma^2
\!+\!\sigma^4\!-\!|H_\mathrm{ue}|^2\right)}
{\left[(P_\mathrm{B}\!+\!||\boldsymbol{w}||_2^2)\left(H_\mathrm{u}H_\mathrm{e}
\!+\!(H_\mathrm{u}+H_\mathrm{e})\sigma^2
\!+\!\sigma^4
\!-\!|H_\mathrm{ue}|^2\right)
\!-\!P_\mathrm{B}\sigma^2(H_\mathrm{e}+\sigma^2)\right](H_\mathrm{e}\!+\!\sigma^2)\sigma^2}\right).
		\end{align}
		\hrulefill
	\end{figure*}
\begin{remark}
	The analytical expression derived in (\ref{eq:whole_cc}) reveals that the KGR depends solely on the beamforming $\boldsymbol{w}$ and the channel's statistical information characterized by covariance matrices. Given the quasi-static nature of covariance matrices in dense scattering environments \cite{yang2020asymptotic}, we adopt a practical assumption that these matrices can be effectively estimated from historical channel measurements {leveraging} well-established estimate techniques \cite{neumann2018covariance}. This allows us to concentrate our efforts on developing optimal beamforming and port selection for the FAS-assisted PLKG.
\end{remark}

Thus, the beamforming design {optimization problem under} spatially correlated channel models {can} be formulated as
    \begin{subequations}\label{problem:P2_cor}
    \begin{align}
		{\rm P2}:\max _{\boldsymbol{w}} \
		&R^{\rm cc}_{\rm SK}  \\
		\ \text { s.t. } 
		&\ \text {C1}, \text {C2}. 
	\end{align}  
    \end{subequations}

\subsection{Analysis of FAS-assisted PLKG}
The FAS introduces a transformative approach to PLKG by dynamically optimizing antenna port configurations to exploit spatial correlation {effectively}. This analysis focuses on maximizing the KGR under transmit power constraints and sparse port activation.
{The KGR $R_{\rm SK}$ is quantified as the achievable secret bits per channel coherence interval, which can be lower bounded as follows\cite{ARCI93,RNDH21}}:
\begin{align}\label{KGR_lower}
    R_{\rm SK} \geq \mathcal{I}\left(\hat{h}_{\rm a};\hat{ h}_{\rm b}\right) -\min\left[\mathcal{I}\left(\hat{h}_{\rm a};\hat{ h}^{\rm d}_{\rm e}\right),\mathcal{I}\left(\hat{h}_{\rm a};\hat{ h}^{\rm u}_{\rm e}\right)\right].
\end{align}
{

In i.i.d. scenario, the mutual information between eavesdropping and legitimate nodes is nearly zero\cite{RNDH21}. {Then, based} on \eqref{KGR_lower}, {we have}

\begin{align}\label{KGR_lower_iid}
    R^{\rm iid}_{\rm SK} \geq \mathcal{I}\left(\hat{h}_{\rm a};\hat{ h}_{\rm b}\right).
\end{align}

In spatially correlated eavesdropping scenario, the Eve's downlink estimates are correlated with those of legitimate nodes. {Then, we have}

\begin{align}\label{KGR_lower_cc}
    R^{\rm cc}_{\rm SK} \geq \mathcal{I}\left(\hat{h}_{\rm a};\hat{ h}_{\rm b}\right) -\mathcal{I}\left(\hat{h}_{\rm a};\hat{ h}^{\rm d}_{\rm e}\right).
\end{align}
}

{Note that the} mutual information between Alice and Bob in \eqref{KGR_lower_iid} and \eqref{KGR_lower_cc}, $\mathcal{I}\left(\hat{h}_{\rm a};\hat{ h}_{\rm b}\right)$,  is mainly influenced by $\boldsymbol{w}^{\sf H}\boldsymbol{J}_{\rm A}\boldsymbol{w}$. {Specifically}, the following lemma {{formally characterizes} this dependence and {provides guidance for} the joint design {of} $\boldsymbol{w}$ and FAS port activation}.
\begin{lemma}
    The objective function in \eqref{problem:P1} increases monotonically with $\beta_{\rm ab} \boldsymbol{w}^{\sf H} \boldsymbol{J}_{\rm A} \boldsymbol{w}$.
    \begin{IEEEproof}
	 {Please refer to} Appendix~\ref{sec:lemma2}. 
    \end{IEEEproof}
\end{lemma}

 {For systems where the number of preset ports $M$ is large, i.e., $(M>>N)$ or the normalized distance between ports is sufficiently small}, the spatially correlated matrix at Alice $\boldsymbol{J}_{\rm A}$ is dominated {by its largest} eigenvalues \cite{NWXW24}. {This property enables} approximate optimization {based on} the largest eigenvalue $\lambda_{\text{max}}$ and its corresponding eigenvector $\boldsymbol{u}_{\lambda_{\text{max}}}$. 
{As {the number of} pre-set ports $M$ decreases, the available degrees of freedom for port selection are reduced, thereby exerting a negative influence on the maximum eigenvalue $\lambda_{\text{max}}$. This results in a decrease in the value of $\boldsymbol{w}^{\sf H}\boldsymbol{J}_{\rm A}\boldsymbol{w}$, leading to a reduction in the KGR presented in {both} \eqref{eq:whole} and \eqref{eq:whole_cc}. 
{In contrast, when} $M$ is sufficiently large, an increase in $N$ causes the sparse vector $\boldsymbol{w}$ to leverage more ports to approximate the eigenvectors $\boldsymbol{u}_{\lambda_{\text{max}}}$, thereby resulting in an increase in the value of $\boldsymbol{w}^{\sf H}\boldsymbol{J}_{\rm A}\boldsymbol{w}$. {Consequently,} this enhances the KGR in \eqref{eq:whole} and \eqref{eq:whole_cc}. 
To determine the beamforming vector $\boldsymbol{w}$ for the optimal combination of $N$ activated ports, one can exhaustively search all possible port subsets and solve the optimization problem for each, then select the highest-performing configuration. {However,} this method generally incurs exceedingly high computational complexity. Alternatively, convex relaxation techniques can {provide} a relatively computationally-efficient sparse beamforming vector $\boldsymbol{w}$, {which will be introduced in next section.}

%To obtain the beamforming vector $\boldsymbol{w}$ corresponding to the optimal combination of $N$ activated ports, a methodology can be adopted whereby the whole range of possible combinations of activated ports is traversed. This is then followed by performing beamforming optimization based on the best selection port. An alternative option is to use convex optimization to obtain the sparse beamforming vector $\boldsymbol{w}$. }

\section{Joint port selection and beamforming optimization}
    {In this section, we aim to develop an iterative algorithm to address the nonconvex items of the formulated problems via SCA combined with Cauchy-Schwarz inequality relaxation. Then, a reweighted $\ell_1$-norm method is applied to advocate the sparsity in the solution to the port activation problem.}
\subsection{Optimization for Problem P1}
    Firstly, we optimize the transmit sparse beamforming $\boldsymbol{w}$ to maximize the KGR $R^{\rm iid}_{\rm SK}$ under the i.i.d. scenario. 
    {According to \textbf{Lemma 2}, the objective function {in} problem ${\rm P1}$ increases monotonically with $\boldsymbol{w}^{\sf H}\boldsymbol{J}_{\rm A}\boldsymbol{w}$. Thus, P1 is equivalent to the following problem}
    \begin{subequations}\label{problem:P1_s}
    \begin{align}
		\max _{\boldsymbol{w}} \
		&\boldsymbol{w}^{\sf H} \boldsymbol{J}_{\rm A} \boldsymbol{w}  \\
		\ \text { s.t. } 
		&\ \text {C1}, \text {C2}.
	\end{align}  
    \end{subequations}
    
    By applying the {eigen-decomposition} $\boldsymbol{J}_\mathrm{A}=\boldsymbol{U}_\mathrm{A}\boldsymbol{\Lambda}^{\sf H}\boldsymbol{U}^{\sf H}_\mathrm{A}$ and introducing the slack {optimization} variable $Z$, problem \eqref{problem:P1_s} can be {equivalently} {transformed into}

    \begin{subequations}\label{problem:P1.1}
    \begin{align}
		\max _{\boldsymbol{w},Z} \ 
		& Z \\
		\ \text { s.t. } 
		&\ \text {C1}, \text {C2}, \\
            &\ \text {C3}{:}\,\boldsymbol{w}^{\sf H}\boldsymbol{U}_{\rm A}\boldsymbol{\Lambda}^{\sf H}\boldsymbol{U}_{\rm A}^{\sf H}\boldsymbol{w} \geq Z.
	\end{align}  
    \end{subequations}

However, problem \eqref{problem:P1.1} is still non-convex due to constraints $\text {C2}$ and $\text {C3}$. To tackle $\text{C2}$, the nonconvex $\ell_0$-norm can be effectively replaced by the sparsity promoting convex $\ell_1$-norm constraint\cite{reweighted,2013L1}, {resulting in}
%By using Cauchy-Schwarz inequality based on paper-select-antenna, 
\begin{equation}
 %\text {C2.1}{:}\,   \|w\|_1 \leq \sqrt{NP_A}.
 \text {C2.1}{:}\,   \|\boldsymbol{w}\|_1 \leq N.
\end{equation}

{
To {handle} $\text{C3}$, the {iterative} SCA technique is applied. {In particular, in the $l$-th iteration of the SCA, a convex subset of $\text{C3}$ is given by}

\begin{equation}
     \text {C3.1}{:}\, -\boldsymbol{w}_{l-1}^{\sf H}\boldsymbol{U}_{\rm A}\boldsymbol{\Lambda}^{\sf H}\boldsymbol{U}_{\rm A}^{\sf H}\boldsymbol{w}_{l-1}+2\mathrm{Re}(\boldsymbol{w}_{l-1}^{\sf H}\boldsymbol{U}_{\rm A}\boldsymbol{\Lambda}^{\sf H}\boldsymbol{U}_{\rm A}^{\sf H}\boldsymbol{w}) \geq Z, 
\end{equation}
}
{i.e., $\text{C3.1} \Longrightarrow \text {C3}$, where $\boldsymbol{w}_{l-1}$ is the beamforming vector obtained from the $l-1$-th iteration of the SCA.}

{As such, a lower-bound performance to \eqref{problem:P1.1} can be obtained by solving}
    \begin{subequations}\label{problem:P1.2}
    \begin{align}
		\max _{\boldsymbol{w},Z} \ 
		& Z \\
		\ \text { s.t. } 
		&\ \text {C1}, \text {C2.1}, \text {C3.1}. 
	\end{align}  
    \end{subequations}

\subsection{Optimization for Problem P2}
To tackle problem P2, we first simplify the optimization problem according to \textbf{Lemma 2}. {Without loss of generality}, {the KGR} $R_{\rm SK}^{\rm cc}$ in \eqref{eq:whole_cc} {can be defined as} $R_{\rm SK}^{\rm cc}=\log_2f(x_0)$, {which increases} monotonically for $x_0$, where
    \begin{align}
         x_0=\beta_{\rm ab}\boldsymbol{w}^{\sf H} \boldsymbol{J}_{\rm A} \boldsymbol{w}-\frac{\|\sqrt{\beta_{\rm ab}\beta_{\rm ae}}\boldsymbol{w}^{\sf H} \boldsymbol{J}_{\rm A} \boldsymbol{w}\|^2}{\beta_{\rm ae}\boldsymbol{w}^{\sf H} \boldsymbol{J}_{\rm A} \boldsymbol{w}+\sigma^2}.   
    \end{align}
    
Thus, the original problem ${\rm P2}$ is equivalent to the following problem:

    \begin{subequations}\label{problem:P2_s}
    \begin{align}
		\max _{\boldsymbol{w}} \
		&\beta_{\rm ab}\boldsymbol{w}^{\sf H}\boldsymbol{J}_{\rm A}\boldsymbol{w}-\frac{\|\sqrt{\beta_{\rm ab}\beta_{\rm ae}}\boldsymbol{w}^{\sf H}\boldsymbol{J}_{\rm A}\boldsymbol{w}\|^2}{\beta_{\rm ae}\boldsymbol{w}^{\sf H}\boldsymbol{J}_{\rm A}\boldsymbol{w}+\sigma^2} \label{problem:P2_s_a}  \\
		\ \text { s.t. } 
		&\ \text {C1}, \text {C2}.
	\end{align}  
    \end{subequations}

    To {address} the nonconvex objective function \eqref{problem:P2_s_a}, slack {optimization} variables $Y$, $I$, and $L$ are introduced to {equivalently reformulate} problem \eqref{problem:P2_s}:
    
    \begin{subequations}\label{problem:P2_YIL_s}
    \begin{align}
		\max _{\boldsymbol{w},Y,I,L} \
		&Y-\frac{\|I\|^2}{L} \label{problem:P2_YIL_s_a}  \\
		\ \text { s.t. } 
		&\ \text {C1}, \text {C2},\\
            &\ \text {C4}{:}\, Y \leq \beta_{\rm ab}\boldsymbol{w}^{\sf H} \boldsymbol{J}_{\rm A} \boldsymbol{w},\\
            &\ \text {C5}{:}\, L \leq \beta_{\rm ae}\boldsymbol{w}^{\sf H} \boldsymbol{J}_{\rm A} \boldsymbol{w}+\sigma^2,\\
            &\ \text {C6}{:}\, I \geq \sqrt{\beta_{\rm \rm ab}\beta_{\rm ae}}\boldsymbol{w}^{\sf H} \boldsymbol{J}_{\rm A} \boldsymbol{w}.
	\end{align}  
    \end{subequations}
{

 By applying the definition $\boldsymbol{J}_\mathrm{A}=\boldsymbol{U}_\mathrm{A}{\boldsymbol{\Lambda}}^{\sf H}{\boldsymbol{U}_\mathrm{A}^{\sf H}}$, {we have}
 \begin{subequations}
 \begin{align}
     &\boldsymbol{J}_\mathrm{A}
     =\boldsymbol{U}_{\mathrm{A}}{\boldsymbol{\Lambda}}^{\sf H}{\boldsymbol{U}_{\mathrm{A}}^{\sf H}}
     =\boldsymbol{U}_{\mathrm{A}}{({\boldsymbol{\Lambda}}^{\sf H})}^{\frac{1}{2}}{({\boldsymbol{\Lambda}}^{\sf H})}^{\frac{1}{2}}{\boldsymbol{U}_{\mathrm{A}}^{\sf H}}.
    \end{align} 
 \end{subequations}
 
{Thus,} problem \eqref{problem:P2_YIL_s} can be rewritten by
    \begin{subequations}\label{problem:P2_YIL_s_f}
    \begin{align}
		\max _{\boldsymbol{w},Y,I,L} \
		&Y-\frac{\|I\|^2}{L} \\
		\ \text { s.t. } 
		&\ \text {C1}, \text {C2},\\
            &\ \text {C4}{:}\, Y \leq \beta_{\rm ab}\boldsymbol{w}^{\sf H}\boldsymbol{U}_{\mathrm{A}}{({\boldsymbol{\Lambda}}^{\sf H})}^{\frac{1}{2}}{({\boldsymbol{\Lambda}}^{\sf H})}^{\frac{1}{2}}{\boldsymbol{U}_{\mathrm{A}}^{\sf H}}\boldsymbol{w},\\
            &\ \text {C5}{:}\, L \leq \beta_{\rm ae}\boldsymbol{w}^{\sf H}\boldsymbol{U}_{\mathrm{A}}{({\boldsymbol{\Lambda}}^{\sf H})}^{\frac{1}{2}}{({\boldsymbol{\Lambda}}^{\sf H})}^{\frac{1}{2}}{\boldsymbol{U}_{\mathrm{A}}^{\sf H}}\boldsymbol{w}+\sigma^2,\\
            &\ \text {C6}{:}\, I \geq \beta_{\rm abe}\boldsymbol{w}^{\sf H}\boldsymbol{U}_{\mathrm{A}}{({\boldsymbol{\Lambda}}^{\sf H})}^{\frac{1}{2}}{({\boldsymbol{\Lambda}}^{\sf H})}^{\frac{1}{2}}{\boldsymbol{U}_{\mathrm{A}}^{\sf H}}\boldsymbol{w},
	\end{align}  
    \end{subequations}
}
where $\beta_{\rm abe}=\sqrt{\beta_{\rm ab}\beta_{\rm ae}}$.

{Note that the objective function is now jointly concave with respect to the optimization variables.} However, constraints \text{C4} and \text{C5} are both nonconvex due to right-hand-side (RHS) term.
{{To handle the constraints,} similar to the previous section,} based on the SCA technique, we provide the following lemma to {establish} {a performance} lower bound.
    \begin{lemma}
	According to Cauchy-Schwartz inequality $\boldsymbol{x}\boldsymbol{y}^{\sf H} \leq \|\boldsymbol{x}\|_2\|\boldsymbol{y}^{\sf H}\|_2$, the inequality during $l$-th iteration can be given by
    \begin{align}
        \|\boldsymbol{f}\|_2 \geq \frac{\mathrm{Re}(\boldsymbol{f}\boldsymbol{f}^{\sf H}_{l-1})}{\|\boldsymbol{f}_{l-1}^{\sf H}\|_2}.\label{Cauchy}
    \end{align}
    \end{lemma}
	\begin{IEEEproof}
		%	This proof is similar to Lemma~1 in \cite{22Globecom} and omitted 
		%	due to the space limitation.
		{Please refer to} Appendix~\ref{sec:lemma3}. 
	\end{IEEEproof}
{
By defining $\boldsymbol{f}=\boldsymbol{w}^{\sf H}\boldsymbol{U}_{\rm A}{(\boldsymbol{\Lambda}^{\sf H})}^{\frac{1}{2}}$, and applying the sparsity-promoting convex $\ell_1$-norm to approximate the nonconvex constraints C2 in \eqref{problem:P2_YIL_s_f}, {a suboptimal solution to \eqref{problem:P2_YIL_s_f} can be obtained via solving}
     \begin{subequations}\label{problem:P2_YIL_s_f_lb}
    \begin{align}
		\max _{\boldsymbol{w},Y,I,L} \
		&Y-\frac{\|I\|^2}{L} \\
		\ \text { s.t. } 
		&\ \text {C1}{:}\,||\boldsymbol{w}||_{2}^2 \leq P_\mathrm{A} , \\
		&\ \text {C2.1}{:}\,||\boldsymbol{w}||_{1} \leq N,\\
            &\ \text {C4.1}{:}\, Y \leq \frac{\beta_{\rm ab}\mathrm{Re}(\boldsymbol{f}{\boldsymbol{f}^{\sf H}_{l-1}})}{\|{\boldsymbol{f}_{l-1}^{\sf H}}\|_2},\\
            &\ \text {C5.1}{:}\, L \leq \frac{\beta_{\rm ae}\mathrm{Re}({\boldsymbol{f}\boldsymbol{f}_{l-1}^{\sf H})}}{\|{\boldsymbol{f}_{l-1}^{\sf H}}\|_2}+\sigma^2,\\
            &\ \text {C6}{:}\, I \geq \beta_{\rm abe}\boldsymbol{f}{\boldsymbol{f}^{\sf H}}.
	\end{align}  
    \end{subequations}
}
\subsection{{Overall} Algorithm}
 An iterative algorithm is provided to obtain an approximate solution for P1 and P2, respectively.
 	\begin{algorithm}[h]
		\caption{Iterative Algorithm for Solving P1 or P2}
		\label{alg:2}
		\begin{algorithmic}[1]
			\Require
			Threshold $\varepsilon_{0}$ and 
			covariance matrix $\boldsymbol{J}_\mathrm{A}$.
			\State 
			Set {iteration index}: $l=0$.
			\State Initial:  ${\boldsymbol{w}}^{(0)}$.
			\Repeat  
			\State For solving P1: 
            \State Update ${\boldsymbol{w}}^{(l+1)}$ by solving Problem  \eqref{problem:P1.2}.
            \State Calculate the objective value $R^{(l)}$ of Problem \eqref{eq:whole}.
                \State For solving P2: 
                \State Update ${\boldsymbol{w}}^{(l+1)}$ by solving Problem  \eqref{problem:P2_YIL_s_f_lb}. 
			\State Calculate the objective value $R^{(l)}$ of Problem \eqref{eq:whole_cc}.
			\State $l \leftarrow l+1$.
			\Until{$|R^{(l)} - R^{(l-1)}|\le \varepsilon_{0}$.}
		\end{algorithmic}
	\end{algorithm} 

    To attain the required number of iterations $\|\boldsymbol{w}\|_0=N$, we develop an iterative reweighted $\ell_1$-norm {approach for} constraint ${\rm C2.1}$, {introducing a diagonal weighting matrix} $\boldsymbol{{V}}$\cite{reweighted}. Especially, {at} the $l$-th iteration, it is denoted as $\boldsymbol{{V}}_{l}=g(\boldsymbol{w}_{l-1})$. To promote sparsity, $g(\boldsymbol{w}_{l-1})$ can be given by
    
\begin{align}\label{regula_pare}
    g(\boldsymbol{w}_{l-1}) \triangleq {\rm{diag}}\left(\frac{1}{|[\boldsymbol{w}_{l-1}]_1|+\gamma},\dots,\frac{1}{|[\boldsymbol{w}_{l-1}]_M|+\gamma}\right),
\end{align}
where $[\boldsymbol{w}_{l-1}]_m$ denotes the $m$-th element of $\boldsymbol{w}$ {at} the $(l-1)$-th iteration {and $\gamma$ is a small regularization parameter to avoid division by zero.} 

\begin{remark}

    The regularization parameter $\gamma$ is crucial for regulating the sparsity of the beamforming vector. {At} the $l$-th iteration, the diagonal {elements of the weighting} matrix $\boldsymbol{{V}}_{l}$ {are} related to the {elements} $[\boldsymbol{w}_{l-1}]_m$. This adaptive scaling maintains numerical stability by stopping weight vector components from growing uncontrollably as $[\boldsymbol{w}_{l-1}]_m$ nears zero. The {parameter} $\gamma$ prevents the iterative optimization from diverging when the components of $\boldsymbol{w}$ approach the null space\cite{CLSelecAntenna13}. 
\end{remark}

Based on constraint $\text{C2.1}$, the weighted $\ell_1$ constraint can be expressed as
\begin{align}
%    ||\boldsymbol{{V}}^{(r)}\boldsymbol{w}||_{1} \leq \sqrt{NP_\mathrm{A}}
    ||\boldsymbol{V}_l\boldsymbol{w}||_{1} \leq N.
\end{align}

The reweighted $\ell_1$ norm iterative algorithm {is summarized as follows}
 	\begin{algorithm}[H]
		\caption{The Reweighted Algorithm for Solving P1 or P2}
		\label{alg:3}
		\begin{algorithmic}[1]
			\Require
			Threshold $\varepsilon_{0}$ and 
			covariance matrix $\boldsymbol{J}_\mathrm{A}$.
			\State 
			Set: $r=0$.
			\State Initial:  ${\boldsymbol{w}}^{(0)}$ and $\boldsymbol{{V}}=\boldsymbol{I}_{\rm M}$.
			\Repeat  
			\State Solve problem P1 or P2 with a weighted $\ell_1$ constraint {adopting} \textbf{{\rm Algorithm 1}}.
			\State Calculate the objective value $\boldsymbol{{V}}^{(r+1)}=g(\boldsymbol{w}^{(r)})$.
			\State $r \leftarrow r+1$.
			\Until{$|R^r - R^{(r-1)}|\le \varepsilon_{0}$.}
		\end{algorithmic}
	\end{algorithm} 

    {The {computational} complexity of the proposed reweighted algorithm is $\mathcal{O}(M^3+M^2N+M)$, which can be attributed to three principal components, including eigen-decomposition with a complexity of $\mathcal{O}(M^3)$, SCA iterations with a complexity of $\mathcal{O}(M^2N)$ and the reweighting computation with a complexity of $\mathcal{O}(M)$.}

\section{Sliding Window-based Port Selection}
In Section IV, {a} convex relaxation technique was employed to derive the computationally efficient sparse beamforming vector ${\boldsymbol{w}}$. {Alternatively}, it is also {possible} to analyze the FAS-assisted PLKG and develop a low-complexity heuristic algorithm {that serves as a practical substitute for} exhaustive searching {methods}, which {typically} exhibits high computational complexity.} According to \textbf{Lemma 2}, the mutual information between Alice and Bob, $\mathcal{I}\left(\hat{h}_{\rm a};\hat{ h}_{\rm b}\right)$, in \eqref{KGR_lower_iid} and \eqref{KGR_lower_cc}, is influenced by the largest eigenvalue, $\lambda_{\max}$. In this section, {leveraging} Rayleigh-quotient theory, we provide the analysis of the largest eigenvalue, which is influenced by zero-order Bessel function of the first kind. {Based on} the analysis, a sliding window-based port selection is proposed to obtain the activated port selection for FAS-assisted PLKG. Firstly, the following discussion about properties of the zero-order Bessel function of the first kind is conducted:

\begin{remark}
The zero-order Bessel function of the first kind expressed as:
$J_0(x) = \sum_{k=0}^{\infty} \frac{(-1)^k}{(k!)^2} \left(\frac{x}{2}\right)^{2k}$,
{is an oscillatory decaying function within the interval $[-1, 1]$ as $x$ increases,} with gradually decreasing amplitude. {This} function reaches its maximum value $J_0(0) = 1$ at $x = 0$, and then fluctuates between positive and negative values and approaches 0 as $x$ increases\cite{Jakes}.
\end{remark}

The $(n, m)$-th element of matrix $\boldsymbol{J}_{\rm A}$ is given by\cite{NWXW24}:
\begin{equation}
    \boldsymbol{J}_{\rm A} (n,m) = J_0\left(\frac{2\pi|n - m|}{M - 1}W\right).
\end{equation}
As $W$ increases, the argument $\frac{2\pi|n - m|}{M - 1}W$ increases for $\forall n \neq m$. Due to the oscillatory decaying property of $J_0(x)$ based on Remark 3, the sum of absolute values of the non-diagonal elements of matrix $\boldsymbol{J}_{\rm A}$ gradually decreases. 

{Based on the Rayleigh-quotient theory, the eigenvalues of the symmetric matrix $\boldsymbol{J}_{\rm A}$ satisfy the following properties:}

\begin{remark}
For a non-zero $M$-dimensional vector $\mathbf{x}$, the Rayleigh-quotient of matrix $\boldsymbol{J}_{\rm A} \in \mathbb{C}^{M\times M}$, is defined as:
  $R(\mathbf{x}) = \frac{\mathbf{x}^{\sf H} \boldsymbol{J}_{\rm A}\mathbf{x}}{\mathbf{x}^{\sf H}\mathbf{x}}$.
  The maximum eigenvalue $\lambda_{\rm{max}}$ and minimum eigenvalue $\lambda_{\rm{min}}$ of matrix $\boldsymbol{J}_{\rm A}$ satisfy:
  $\lambda_{\rm{min}} \leq R(\mathbf{x}) \leq \lambda_{\rm{max}}$
\end{remark}

As \(W\) increases, the sum of absolute values of the non-diagonal elements of matrix $\boldsymbol{J}_{\rm A}$ decreases. 
%These non-diagonal elements reflect the coupling degree between different rows (columns) of the matrix. The reduction in the sum of absolute values weakens the interaction between rows (columns). 
From the perspective of the Rayleigh-quotient, for any vector $\mathbf{x}$, we have:

\begin{equation}
    \mathbf{x}^{\sf H} \boldsymbol{J}_{\rm A}\mathbf{x} = \sum_{n=1}^{M}\sum_{m=1}^{M} x_n J_{n,m} x_m.
\end{equation}

\begin{figure}[H]
		\centering
		\includegraphics[width=2.7in]{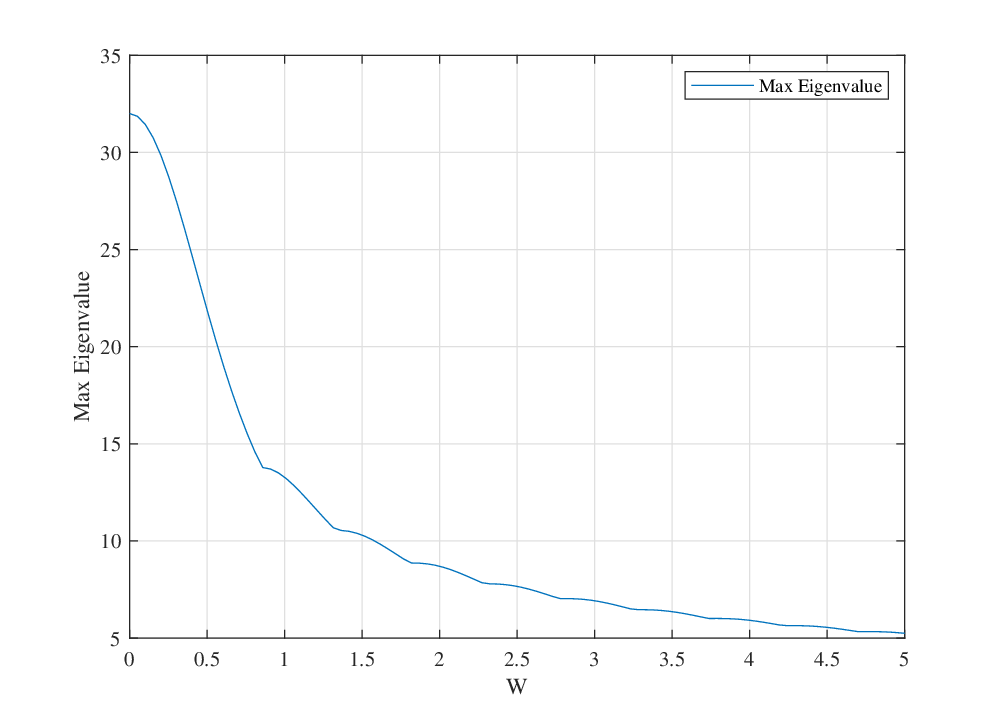}
		\caption{The maximum eigenvalue of matrix $\boldsymbol{J}_{\rm A}$ versus $W$.}
		\label{JA_W}
	\end{figure}

The diagonal elements of matrix $\boldsymbol{J}_{\rm A}$ are $J_{n,n} = J_0(0) = 1$, which are independent of $W$. The decrease in {magnitude of the} non-diagonal elements leads to a relatively smaller value of $\mathbf{x}^{\sf H} \boldsymbol{J}_{\rm A}\mathbf{x}$, thereby causing the maximum eigenvalue to decline, {as shown} in Fig.~\ref{JA_W}. 
 \begin{comment}   
According to \textbf{Lemma 2}, the decline in the eigenvalue is associated with a reduction in the value of $\boldsymbol{w}^{\rm T} \boldsymbol{J}_{\rm A} \boldsymbol{w}$, which exerts a further influence on the decline in the KGR of FAS-assisted communication.

 Considering that closer distances between $N$ activated ports are preferable, a sliding window $\boldsymbol{S}$ of size $N$ is applied to the 1-dimensional $M$ ports to select $N$ consecutive elements as initial values. Specifically:
\end{comment}

{\textbf{Lemma 2} suggests that a decrease in $\lambda_{\max}$ directly reduces $\boldsymbol{w}^{\sf H}\boldsymbol{J}_{\rm A}\boldsymbol{w}$ and consequently lowers the KGR of FAS-assisted communications. To counteract this effect by {exploiting} closely spaced active ports, we {propose} a sliding-window initialization over the one-dimensional array. The key steps {of this method} {are} summarized in the following:}

\emph{Step 1: Candidate Window Sets Generation.} Generate $M - N + 1$ candidate window sets by sliding a window $\boldsymbol{S}$ of length $N$ across the $M$ ports. 

\emph{Step 2: Sliding Window Selection.} Compute $\boldsymbol{w}^{\sf H} \boldsymbol{J}_{\rm A}\boldsymbol{w}$ for each window set and select the window set with the maximum value {to serve as} the initial beamforming vector $\boldsymbol{w}^{(0)}$.

\emph{Step 3: Power Optimization.} Perform power optimization on the beamforming vector $\boldsymbol{w}^{(0)}$ within the selected window set to satisfy the transmit power constraint.

The proposed window selection process efficiently reduces the initialization overhead while leveraging the spatial correlation property of the FAS. Especially, compared to {exhaustively} traversing all possible port {subset} combinations (e.g., the number of combinations is more than 200,000 for $M=32, N=5$), the computational complexity is reduced {significantly} from exponential to polynomial time $\mathcal{O}((M+N-1)M^2)$. This is analogous to the reweighted method proposed in Section IV. C. { The sparse vector and transformation matrix computation of the sliding window selection-based algorithm is summarized as follows}

\begin{algorithm}[H]
    \caption{Sparse Vector and Transformation Matrix {Computation}}
    \label{alg:eigen-transformation}
    \begin{algorithmic}[1]
        \Require
        symmetric matrix $\boldsymbol{J}_{\rm A} \in \mathbb{R}^{M \times M}$, selection dimension $N$, Power budget $P_{\rm A}$.
        \State Compute eigenvalues and eigenvectors:
        $[\boldsymbol{V}, \boldsymbol{D}]$.
        \State
        $\lambda_{\text{max}} = \max(\text{diag}(\boldsymbol{D}))$, {index for maximum value $\text{I} = \text{argmax}(\text{diag}(\boldsymbol{D}))$}.
        \State Extract the eigenvector corresponding to the maximum eigenvalue:
        $\boldsymbol{u} = \boldsymbol{V}( \text{I})$.
        \State
        Initialize $\boldsymbol{w}$, maximum value {$F_{\max}$}.
    \For{$s = 1$ to $M - N + 1$} 
            \State Determine the index set $\mathcal{I}$ covered by the current sliding window:
            $\mathcal{I} \gets [s, s + 1, \cdots, s + N - 1]$.
            \State Initialize the current sparse vector $\boldsymbol{w}$ as a zero vector:
            $\boldsymbol{w}^{\rm ini} = \boldsymbol{0}_M$.
            \State Assign the values of $\boldsymbol{u}$ at the corresponding indices to $\boldsymbol{w}$:
            $\boldsymbol{w}(\mathcal{I}) = \boldsymbol{u}(\mathcal{I})$.
            \State
            {Calculate $F_{\rm cur}$} 
            \If{{$F_{\rm cur}$ $>$ $F_{\max}$}} 
                \State Update:
                {$F_{\max}$ $=$ $F_{\rm cur}$}, 
                $\boldsymbol{w}^{\rm ini} = \boldsymbol{w}$,
                $\mathcal{I}^* = \mathcal{I}$, 
            \EndIf
        \EndFor
        \State Normalized $\boldsymbol{w}^{\rm ini}$ with power budget $P_{\rm A}$.
        \State Construct the transformation matrix:
        $\boldsymbol{S} = \boldsymbol{0}_{M \times M},\ \boldsymbol{S}(\mathbf{i}^{\rm ini}, \mathbf{i}^{\rm ini}) = \boldsymbol{I}_N$.
        \State
        \textbf{Output}: $\boldsymbol{w}^{\rm ini},\ \boldsymbol{S}$.
    \end{algorithmic}
\end{algorithm}

Defining $\boldsymbol{\bar{w}}=\boldsymbol{S}\boldsymbol{w}$, the sliding window selection-based iterative algorithm can be given by
 	\begin{algorithm}[H]
		\caption{The Sliding Window Selection-based Algorithm}
		\label{alg:4}
		\begin{algorithmic}[1]
			\Require
			Threshold $\varepsilon_{0}$, 
			covariance matrix $\boldsymbol{J}_\mathrm{A}$, the number of activated port $N$ and power budget $P_{\rm A}$.
			\State 
			Set: $r=0$.
			\State Initial:  Use \textbf{{\rm Algorithm 3}} to {obtain} initial beamforming vector $\boldsymbol{w}^{\rm ini}$ and sliding window $\boldsymbol{S}$.
			\Repeat
                \State For solving P1: 
                \State Update ${\boldsymbol{\bar{w}}}^{(l+1)}$ by solving Problem  \eqref{problem:P1.2}.
                \State Calculate the objective value $R^{(l)}$ of Problem \eqref{eq:whole}.
                \State For solving P2: 
                \State Update ${\boldsymbol{\bar{w}}}^{(l+1)}$ by solving Problem  \eqref{problem:P2_YIL_s_f_lb}. 
			\State Calculate the objective value $R^{(l)}$ of Problem \eqref{eq:whole_cc}.
			%\State Update ${\boldsymbol{\bar{w}}}^{(l+1)}$ by solving Problem \eqref{problem:P1.2} or \eqref{problem:P2_YIL_s_f_lb}. 
			%\State Calculate the objective value $R^{(l)}$ of Problem \eqref{eq:whole} or \eqref{eq:whole_cc}.
			\State $l \leftarrow l+1$.
			\Until{$|R^{(l)} - R^{(l-1)}|\le \varepsilon_{0}$.}
		\end{algorithmic}
	\end{algorithm}

	\section{Simulation Results}\label{sec:simulation}
	In this {section}, we provide simulation results to illustrate the PLKG performance of the proposed method and {investigate the} impact of spatial correlation on {the} KGR. 
	
\subsection{Simulation Settings}
Specific parameters are detailed in Table \ref{simulation parameter}. In the simulation, Alice and Bob are positioned at coordinates $(0 ~\rm{m}, 0~\rm{m})$ and $(70~\rm{m}, 0~\rm{m})$, respectively. In the spatially correlated eavesdropping scenario, Eve’s location is randomly distributed within a circle centered at Bob, whereas in the independent and identically distributed scenario, Eve’s location is randomly distributed in {distinct} regions to ensure statistical independence of eavesdropping channels.
\begin{table}[h]
    \centering
    \caption{Parameter Table}
    \label{simulation parameter}
    \begin{tabular}{|c|c|c|}
        \hline
        Parameter & Meaning & Value \\
        \hline
        $M$ & The number of FAS pre-set port & $32$\cite{NWXW24} \\
        \hline
        $W$ & Normalized distance of FAS & $0.5\lambda$\cite{NWXW24} \\
        \hline
        $N$ & The number of FAS active antenna & $5$\cite{NWXW24} \\
        \hline
        $\sigma^2$ & Noise variance & $-80$ dBm\cite{HLQH24} \\
        \hline
        $P_{\rm A}$ & Power budget of Alice & $20$ dBm\cite{HLQH24} \\
        \hline
        $P_{\rm B}$ & Power budget of Bob & $20$ dBm\cite{HLQH24} \\
        \hline
        $\gamma_0$ & Path loss at $1$ m & $-30$ dB\cite{HLQH24} \\
        \hline
        $\alpha_0$ & Path loss exponent & $2$\cite{HLQH24} \\
        \hline
        $\epsilon$ & Convergence threshold & $10^{-4}$\cite{HLQH24} \\
        \hline
    \end{tabular}
\end{table}

The considered algorithms are defined as follows:
\begin{itemize}
    \item \textbf{``FA Opt"}: In this case, Alice is equipped with conventional fixed position antennas, and the beamforming is optimized with the iterative algorithm $\mathbf{Algorithm~1}$.
    
    \item \textbf{``FA MRC"}: In this case, Alice is equipped with conventional fixed-position antennas, and the beamforming is optimized with {the} MRC technique {as} in \cite{NWXT24,VUMG24,WSTZ21,RWJK24}.

    \item \textbf{``FAS Opt"}: In this case, Alice is equipped with {an} FAS, and the beamforming is optimized with the iterative algorithm $\mathbf{Algorithm~1}$ while activated ports are given by the reweighted algorithm $\mathbf{Algorithm~2}$.
    
    \item {\textbf{``FAS Traverse"}: In this case, Alice is equipped with FAS, and the activated port combination is obtained by traversing all possible combinations \cite{CCXYW25}.}

    \item \textbf{``FAS Sliding Window"}: In this case, Alice is equipped with FAS. The optimal beamforming and activated ports are obtained by using $\mathbf{Algorithm~4}$.

    \item \textbf{``FAS Sliding Window/no"}: In this case, Alice is equipped with FAS, and only activated ports are obtained by exploiting $\mathbf{Algorithm~3}$.
\end{itemize}
	
	\subsection{Impact of the Power Budget}
     \begin{figure}[h]
		\centering
		\includegraphics[width=2.7in]{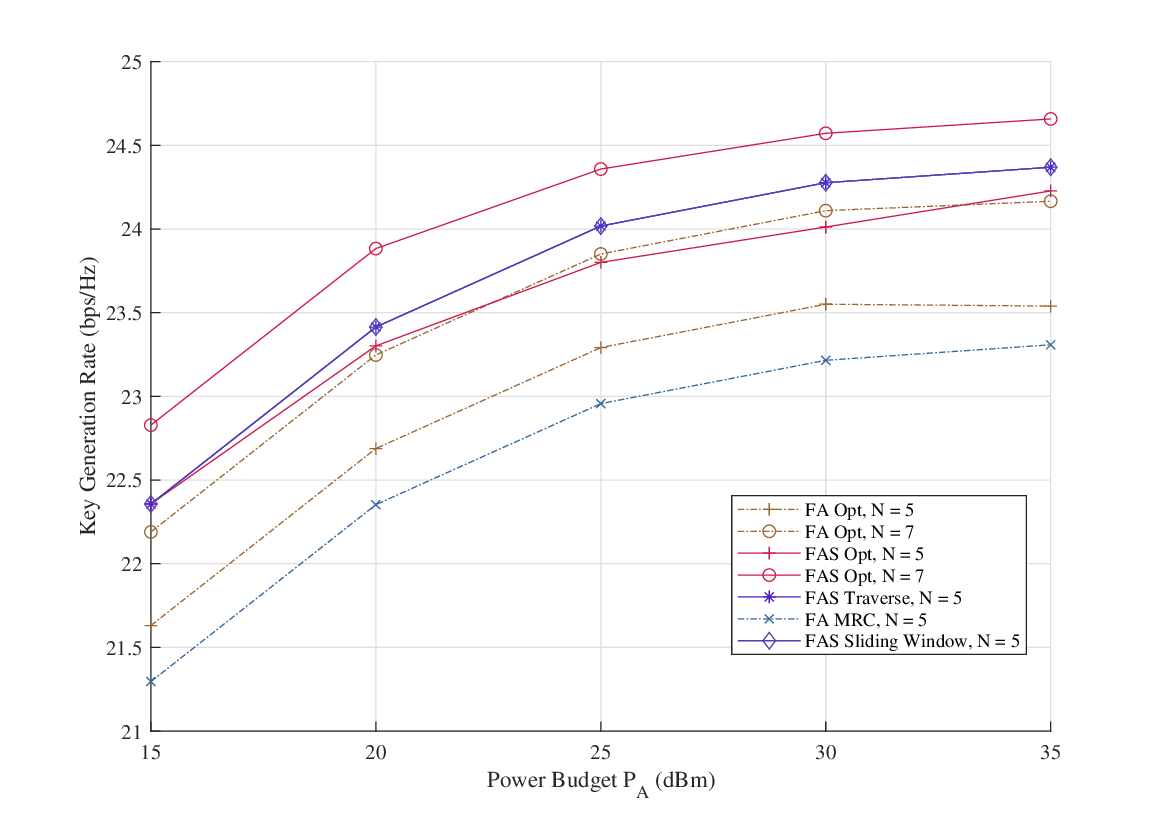}
		\caption{Key generation rate $R^{\rm iid}_{\rm SK}$ versus power budget $P_{\rm A}$ in i.i.d. channel with different schemes between FA and FAS.}
		\label{iid_Rsk_PA_Scheme}
	\end{figure}

        \begin{figure}[h]
		\centering
		\includegraphics[width=2.7in]{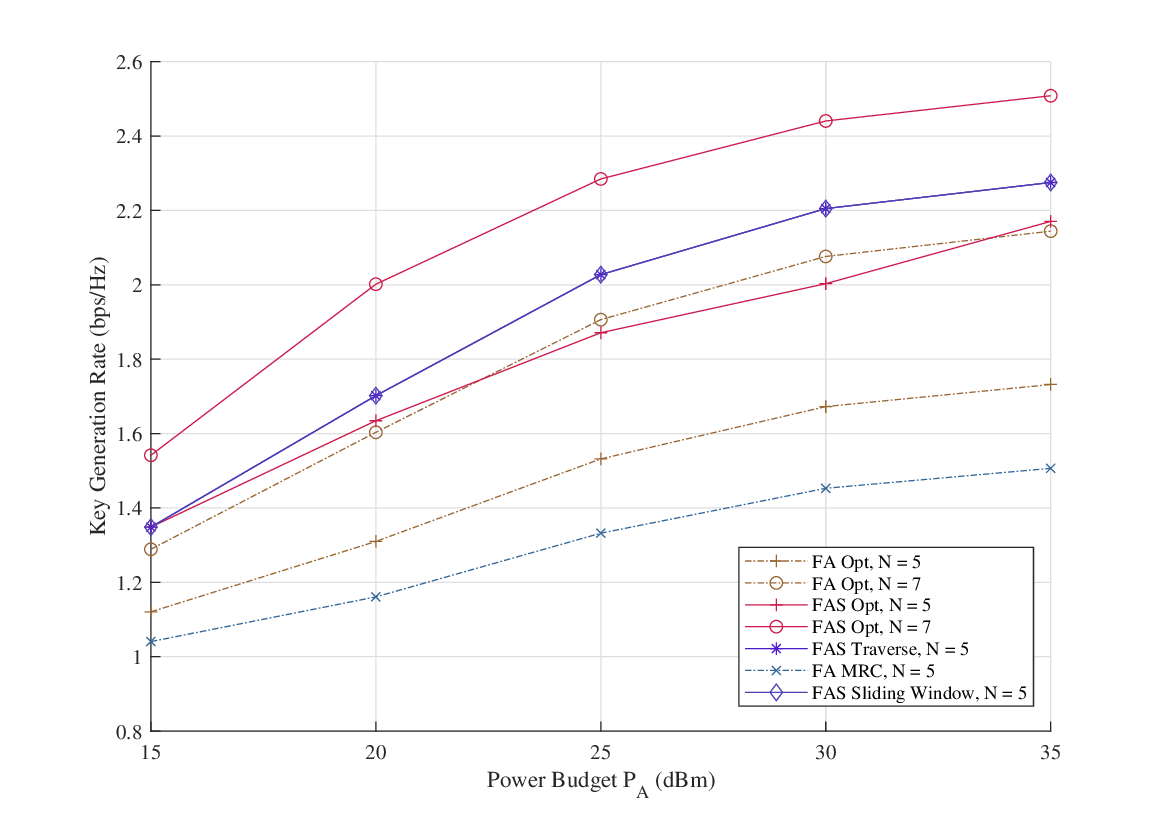}
		\caption{Key generation rate $R^{\rm cc}_{\rm SK}$ versus power budget $P_{\rm A}$ in spatially correlated channel with different schemes between FA and FAS.}
		\label{cc_Rsk_PA_Scheme}
	\end{figure}
    
        As demonstrated in Fig.~\ref{iid_Rsk_PA_Scheme} and Fig.~\ref{cc_Rsk_PA_Scheme}, the KGR of all schemes exhibits a monotonically increasing trend with the transmit power budget $P_{\rm A}$ {in both independent and spatially correlated channel scenarios}. This behavior aligns with the theoretical analysis {presented} in \textbf{Lemma~1} and \textbf{Lemma~2}, which {link the KGR to} mutual information {that depends on} $\boldsymbol{w}^{\sf H}\boldsymbol{J}_{\rm A}\boldsymbol{w}$ and transmit power budget. 
        Notably, as $P_{\rm A}$ increases, the KGR growth rate gradually slows and approaches saturation. This saturation is theoretically underpinned by \textbf{Lemma~4}, which proves that the derivative of the KGR with respect to $P_{\rm A}$ diminishes to zero as $P_{\rm A}$ {becomes} sufficiently large. {Therefore, unilaterally increasing Alice's power budget $P_{\rm A}$ to a sufficiently large extent cannot lead to significant improvement in the KGR, but instead increases the energy consumption of hardware devices.}
        
        {Comparing} the KGR between FAS and FA, it can be observed that the FAS-based schemes, including ``FAS Opt", ``FAS Traverse", and ``FAS Sliding Window", have significantly higher KGR than the FA-based schemes. It is noteworthy that the FAS scheme achieves similar or better performance with fewer activated antennas (``FAS Opt, $N = 5$") than the FA scheme with more activated antennas (``FA Opt, $N = 7$"). {This superiority stems from dynamic sparse port selection, which concentrates transmit power on spatially correlated ports to maximize $\boldsymbol{w}^{\sf H}\boldsymbol{J}_{\rm A}\boldsymbol{w}$, thereby enhancing the mutual information more efficiently than {a} fixed-antenna {arrangement}. }
    
        \begin{figure}[h]
		\centering
		\includegraphics[width=2.7in]{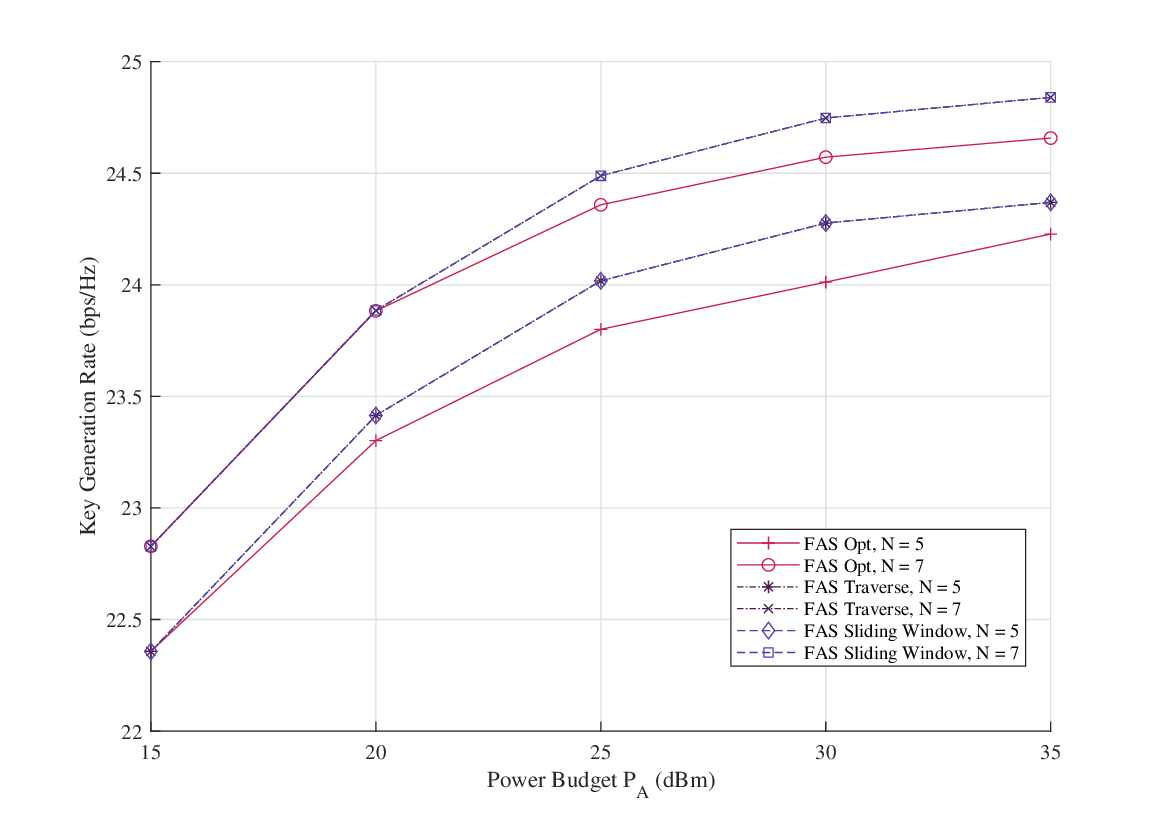}
		\caption{Key generation rate $R^{\rm iid}_{\rm SK}$ versus power budget $P_{\rm A}$ in i.i.d. channel with different schemes by {exploiting} FAS.}
		\label{iid_Rsk_PA_FAS_Scheme}
	\end{figure}
    
        \begin{figure}[h]
		\centering
		\includegraphics[width=2.7in]{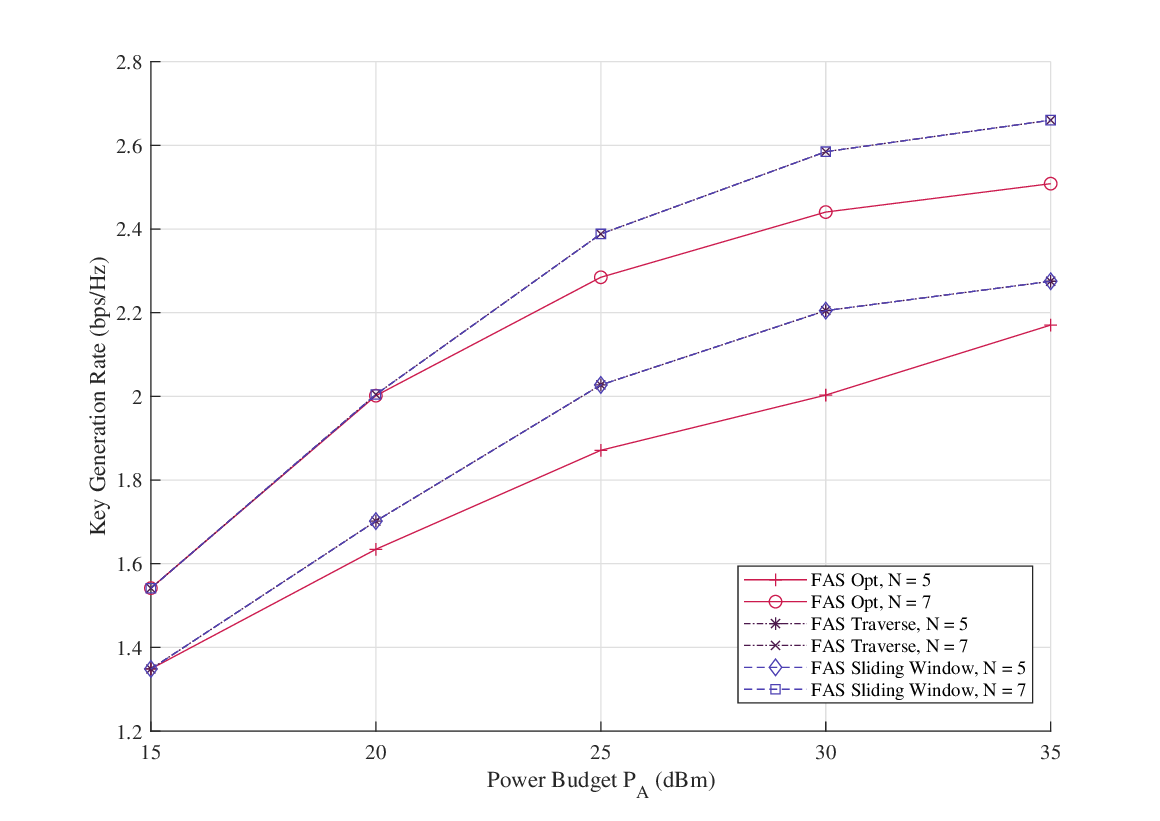}
		\caption{Key generation rate $R^{\rm cc}_{\rm SK}$ versus power budget $P_{\rm A}$ in spatially correlated channel with different schemes by {exploiting} FAS.}
		\label{cc_Rsk_PA_FAS_Scheme}
	\end{figure}

        Furthermore, the comparisons of the KGR between different FAS-based schemes are shown in Fig.~\ref{iid_Rsk_PA_FAS_Scheme} and Fig.~\ref{cc_Rsk_PA_FAS_Scheme}. There is a performance gap between the ``FAS Opt" scheme based on the reweighted $\ell_1$-norm optimization algorithm and ``FAS Traverse" by traversing all possible port combinations. {{This gap arises} because the regularization parameter in \eqref{regula_pare} causes the direction of port selection to deviate from the optimal port combination with the increase of the power budget.
        In contrast, the ``FAS Sliding Window" effectively guides the optimization process to approach the optimal port combination by selecting consecutive highly correlated ports as the initial values based on the Rayleigh-quotient analysis and the spatial correlation property of the Bessel function. }{As a result, {it consistently achieves performance levels similar to those of an exhaustive search, but with substantially lower complexity.}}

	\subsection{Impact of RF Chains} 
            \begin{figure}[h]
		\centering
		\includegraphics[width=2.7in]{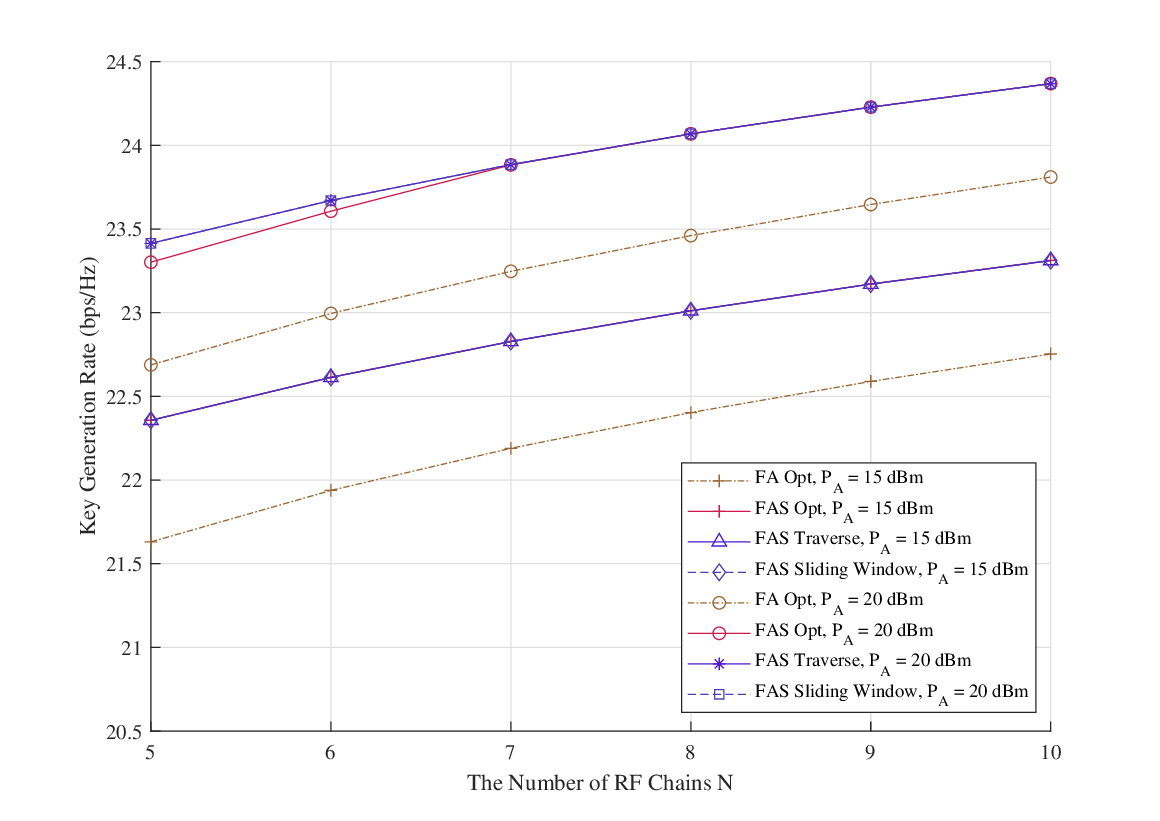}
		\caption{Key generation rate $R^{\rm iid}_{\rm SK}$ versus the number of RF chains $N$ in i.i.d. channel with different power budget $P_{\rm A}$.}
		\label{iid_Rsk_N_P}
	\end{figure}
    
        \begin{figure}[h]
		\centering
		\includegraphics[width=2.7in]{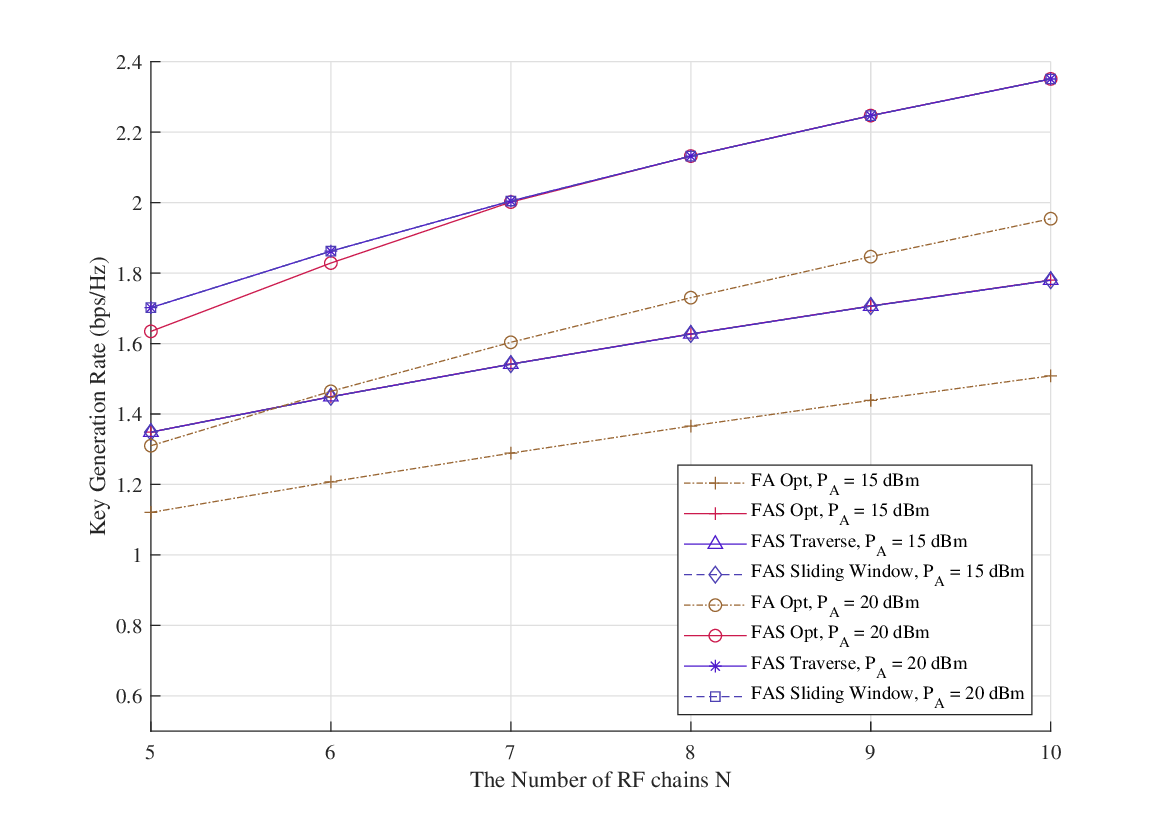}
		\caption{Key generation rate $R^{\rm cc}_{\rm SK}$ versus the number of RF chains $N$ in spatially correlated channel with different power budget $P_{\rm A}$.}
		\label{cc_Rsk_N_P}
	\end{figure}

        Fig.~\ref{iid_Rsk_N_P} and Fig.~\ref{cc_Rsk_N_P} illustrate that the KGR increases monotonically with the number of activated antennas $N$ in both channel scenarios, a direct consequence of extended spatial degrees of freedom. {{As $N$ grows, the sparse beamforming vector can more accurately }approximate the eigenvector corresponding to the largest eigenvalue $\lambda_{\text{max}}(\boldsymbol{J}_{\rm A})$, as discussed in \textbf{Lemma 2} and Rayleigh-quotient analysis. 
        {Indeed, for} the term $\boldsymbol{w}^{\sf H}\boldsymbol{J}_{\rm A}\boldsymbol{w}$ in \eqref{eq:whole} and \eqref{eq:whole_cc},  {an increase in the number of activated ports within fixed {available} pre-set ports yields a corresponding augmentation in the available spatial degrees of freedom {for} the beamforming vector $\boldsymbol{w}$. Consequently, this facilitates a more efficient implementation of channel spatial correlation, which aligns with the predominant eigenmode of the spatial correlation matrix $\boldsymbol{J}_{\rm A}$, subsequently maximizing the mutual information between legitimate channels.}

        {Notably, {comparing the growth rates of KGR} with respect to $N$ in both scenarios, as demonstrated in Fig.~\ref{iid_Rsk_N_P} and Fig.~\ref{cc_Rsk_N_P}, reveals that KGR increases by 38.2\% in the spatially correlated scenario and 4.2\% in the i.i.d. scenario, indicating {a} significantly higher sensitivity of KGR to the number of RF chains in the former. This disparity arises because the KGR of the i.i.d. scenario, $R^{\rm iid}_{\rm SK}$, is independent of Eve’s channel and its growth depends only on $\boldsymbol{w}^{\sf H} \boldsymbol{J}_{\rm A} \boldsymbol{w}$, {whereas} the KGR of the spatially correlated scenario, $R_{\rm SK}^{\rm cc}$, relies on conditional mutual information that accounts for Eve's channel. {Thus, activating more} ports {improves} the KGR by both strengthening the legitimate mutual information of the channel via increasing \(\boldsymbol{w}^{\sf H} \boldsymbol{J}_{\rm A} \boldsymbol{w}\) and mitigating the part related to Eve by reducing $\frac{|\sqrt{\beta_{\rm ab}\beta_{\rm ae}}\boldsymbol{w}^{\sf H}\boldsymbol{J}_{\rm A}\boldsymbol{w}|^2}{\beta_{\rm ae}\boldsymbol{w}^{\sf H}\boldsymbol{J}_{\rm A}\boldsymbol{w}+\sigma^2}$.}

        {Furthermore, as} demonstrated in Fig.~\ref{iid_Rsk_N_P} and Fig.~\ref{cc_Rsk_N_P}, a discrepancy in performance emerges between the ``FAS Opt" and ``FAS Traverse" methods for $N < 7$. {This} discrepancy arises because the initial random port activation and regularization parameter in \eqref{regula_pare} of the reweighted $\ell_1$-norm algorithm deviates from the optimal port combination with fewer activated ports. In contrast, the ``FAS Sliding Window" scheme initially selects highly correlated consecutive ports by leveraging the window function $\boldsymbol{S}$ and optimizes the beamforming vector $\boldsymbol{w}$ without the regularization parameter, thus improving the KGR.
        }
        
        \begin{figure}[h]
		\centering
		\includegraphics[width=2.7in]{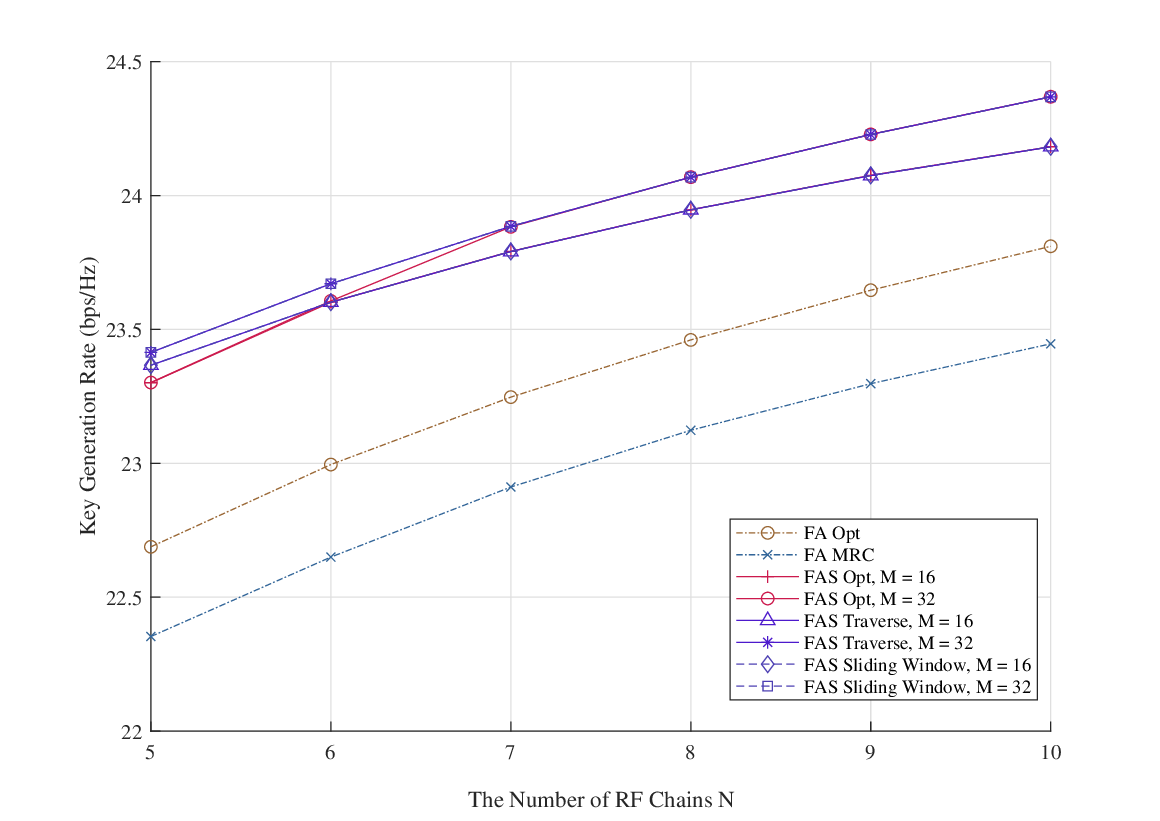}
		\caption{Key generation rate $R^{\rm iid}_{\rm SK}$ versus the number of RF chains $N$ in i.i.d. channel with different pre-set ports $M$.}
		\label{iid_Rsk_N_scheme}
	\end{figure}
    
       \begin{figure}[h]
		\centering
		\includegraphics[width=2.7in]{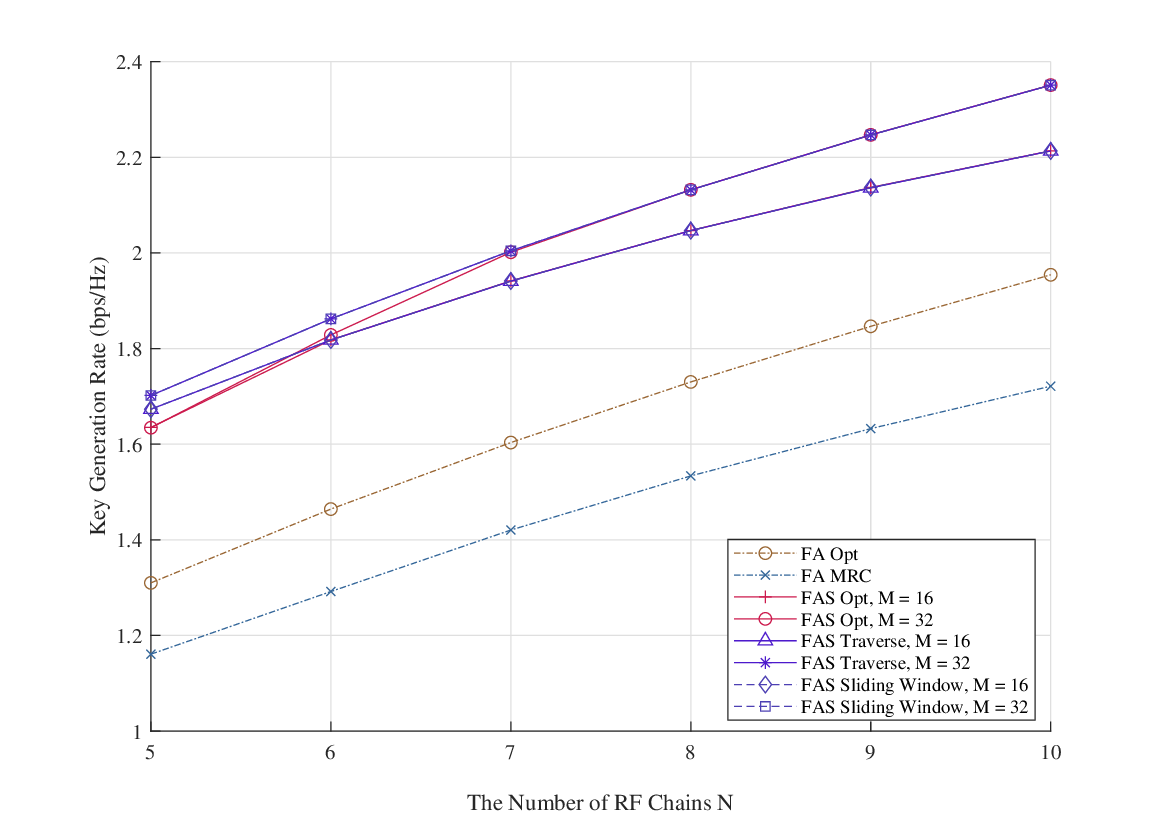}
		\caption{Key generation rate $R^{\rm cc}_{\rm SK}$ versus the number of RF chains $N$ in spatially correlated channel with different pre-set ports $M$.}
		\label{cc_Rsk_N_scheme}
	\end{figure}

    In addition, the number of pre-set ports $M$ also influences the KGR of FAS as shown in Fig.~\ref{iid_Rsk_N_scheme} and Fig.~\ref{cc_Rsk_N_scheme}. This {highlights} that {increasing the number of} activated antennas {enhances} the mutual information of the legitimate channel by extending the spatial degrees of freedom. The FAS-based schemes {exhibit consistently} higher KGR than the FA-based schemes. {{In fact,} the decrease of the number of FAS preset ports $M$ limits the exploitation of spatial correlation by FAS. When $M=16$, the KGR of ``FAS Opt" for $N=10$ in the independent channel decreases compared to $M=32$. 
    {This occurs because {a smaller} $M$ shrinks the available subset of highly correlated ports, limiting the potential to maximize the mutual information of the channel {through the term} $\boldsymbol{w}^{\sf H} \boldsymbol{J}_{\rm A} \boldsymbol{w}$ in \eqref{eq:whole}. Therefore, when $M$ is not sufficiently large,} the KGR is limited by the number of preset ports $M$, {which aligns with the analysis in} Section III. C.
    }

        \subsection{Impact of Normalized Size} 
        \begin{figure}[h]
		\centering
		\includegraphics[width=2.7in]{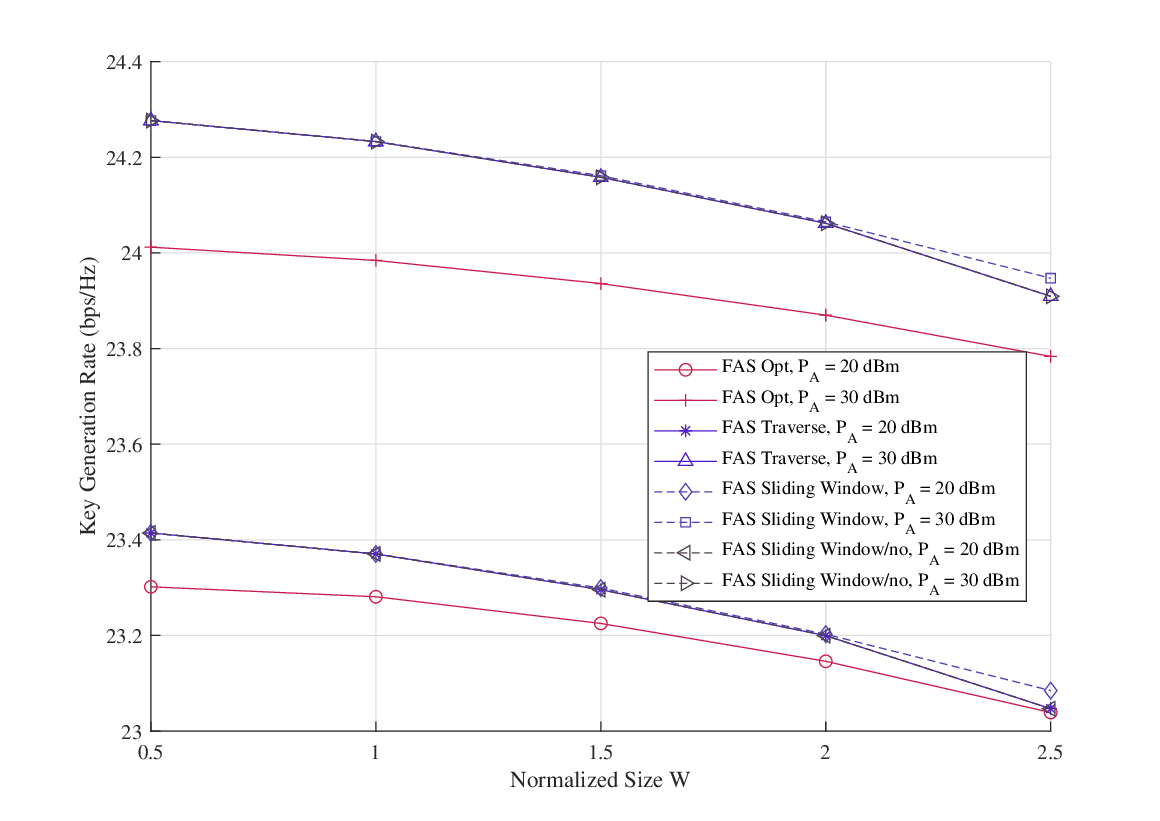}
		\caption{Key generation rate $R^{\rm iid}_{\rm SK}$ versus normalized size of FAS $W$ in i.i.d. channel with different schemes.}
		\label{iid_Rsk_W_Scheme}
	\end{figure}
        
        \begin{figure}[h]
		\centering
		\includegraphics[width=2.7in]{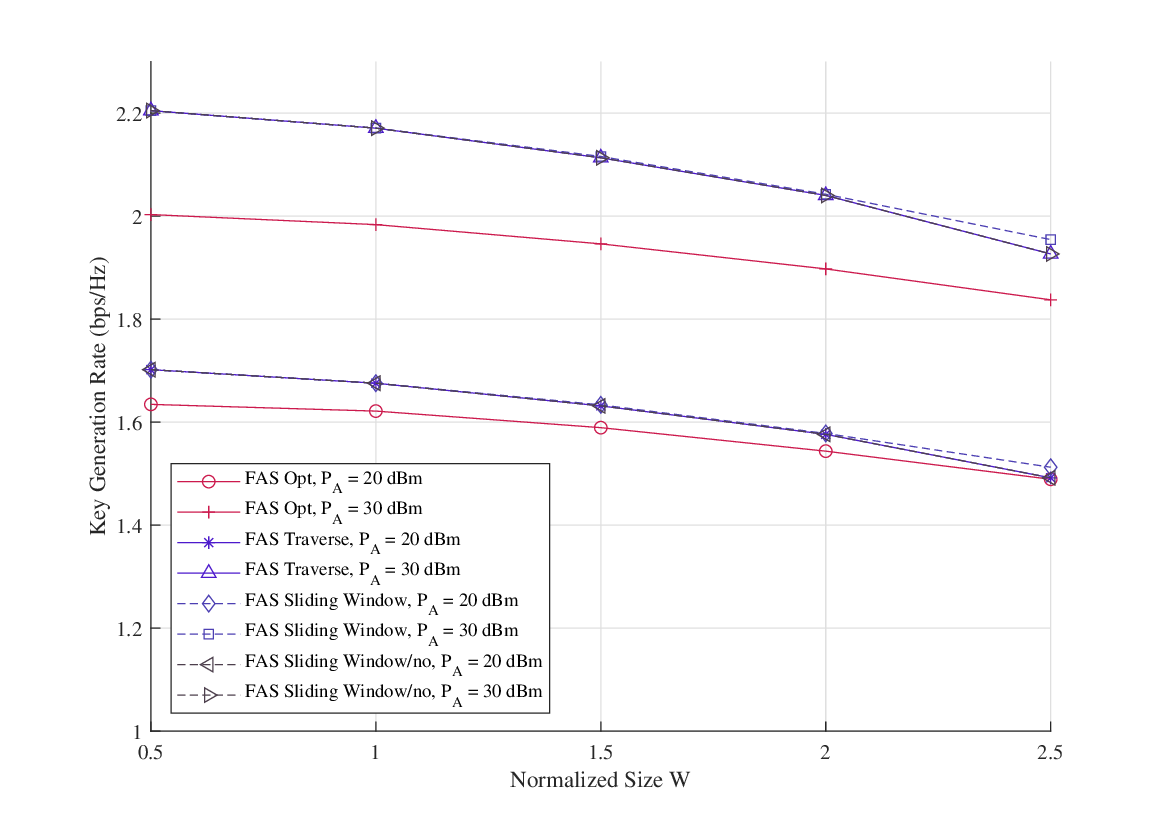}
		\caption{Key generation rate $R^{\rm cc}_{\rm SK}$ versus normalized size of FAS $W$ in spatially correlated channel with different schemes.}
		\label{cc_Rsk_W_Scheme}
	\end{figure}

    {In Fig.~\ref{iid_Rsk_W_Scheme} and Fig.~\ref{cc_Rsk_W_Scheme}, as the normalized size $W$ of the FAS increases, the KGR decreases monotonically, primarily due to {the} reduced spatial correlation between activated ports. The reason is that the KGR is monotonically increasing with $\boldsymbol{w}^{\sf H}\boldsymbol{J}_{\rm A}\boldsymbol{w}$, which is directly influenced by $\lambda_{\max}(\boldsymbol{J}_{\rm A})$ since $\boldsymbol{w}^{\sf H}\boldsymbol{J}_{\rm A}\boldsymbol{w} \leq \lambda_{\max}(\boldsymbol{J}_{\rm A})\|\boldsymbol{w}\|_2^2$ according to Rayleigh-quotient theory. Thus, a smaller $\lambda_{\max}(\boldsymbol{J}_{\rm A})$ under larger $W$ reduces $\boldsymbol{w}^{\sf H}\boldsymbol{J}_{\rm A}\boldsymbol{w}$, leading to diminished mutual information \(\mathcal{I}(\hat{h}_a;\hat{h}_b)\) and lower KGR.  
    It can be observed that “FAS Sliding Window” maintains a higher KGR than “FAS Opt” across different $W$, demonstrating that sliding window-based port selection {can always determine effective activated} port combination such that the KGR is improved, as guided by the Bessel function’s decay characteristics and the Rayleigh-quotient theory, as analyzed in \textbf{Remark 4}. It can be noticed that when $W$ becomes large such as $W=2.5$, the combination of {highly correlated} ports and beam shaping optimization ``FAS Sliding Window" can further increase the KGR compared to simply considering high correlation ports ``FAS Sliding Window/no", because the increase of $W$ results in a decrease in the maximum eigenvalue of $\boldsymbol{J}_{\rm A}$, and the resulting sparse beam shaping vector $\boldsymbol{w}$ based on the corresponding eigenvector $\boldsymbol{u}_{\max}$, {requires} further {optimization} to satisfy the optimal power allocation.}

\section{Conclusion}
    This paper {investigated PLKG} in multi-antenna base station systems by applying the FAS to dynamically optimize radio environments. {We proposed novel} FAS-assisted PLKG models integrating transmit beamforming and sparse port selection under {both} i.i.d. and spatially correlated channels, {and derived} a closed-form KGR expression via reciprocal channel probing while considering eavesdropper channels in correlated scenarios. The {resulted} nonconvex optimization problem for maximizing KGR under power and sparse port constraints {was addressed by exploiting SCA technique, Cauchy-Schwarz inequality, and a reweighted $\ell_1$-norm algorithm.} To improve optimization initialization, a sliding window-based port selection method {was} introduced {utilizing} Rayleigh-quotient theory. Simulations {demonstrated} {that} the FAS-PLKG scheme significantly outperforms traditional FA-PLKG in both channel environments, with the sliding window method achieving higher KGR than the reweighted $\ell_1$-norm approach. {Furthermore, it} was shown that the FAS achieves higher KGR with fewer RF chains through dynamic sparse port selection, which effectively reduces the resource overhead. {Finally}, the sliding window approach proposed in this paper approximates the optimal port selection compared to the reweighted $\ell_1$-norm method, offering valuable insights for practical system deployment.
	
	\appendix
	
	%% appendices
	\begin{appendices}
		\subsection{Proof of Lemma~1}\label{sec:covariance_calculation}
		First, we {derive} the covariance of channel estimate $\hat{ h}_{a}$ as 
		\begin{align}
			\mathcal{R}_{\rm aa} 
			= P_\mathrm{B} \boldsymbol{w}^{\sf H} \mathbb{E}\{\boldsymbol{h}_{\rm ab}  \boldsymbol{h}_{\rm ab}^{\sf H} \} \boldsymbol{w} +  ||\boldsymbol{w}||_{2}^{2} \sigma_{\rm a}^{ 2 }. \label{Ra}
		\end{align}
		By exploiting the channel correlation matrix, the first term of the right-hand side of (\ref{Ra}) is calculated as {$\mathbb{E}\{\boldsymbol{h}_{\rm ab}  \boldsymbol{h}_{\rm ab}^{\sf H} \} = \beta_{\rm  ab } \boldsymbol{J}_{\rm A}$}.

		Similarly, we have
		%the other covariances can be calculated as 
		{\begin{align}
			\mathcal{R}_{\rm bb} &= \beta_{\rm ab} \boldsymbol{w}^{\sf H} \boldsymbol{J}_{\rm A} \boldsymbol{w} +  \sigma_{\rm  b } ^ { 2 }, \\
		\mathcal{R}_{\rm ab}&=\mathcal{R}_{\rm ba}=\sqrt{P_{\mathrm{B}}}\beta_{\rm ab} \boldsymbol{w}^{\sf H} \boldsymbol{J}_{\rm A} \boldsymbol{w}=\mathcal{R}^{\sf H}_{\rm ba},\\
            \mathcal{R}_{\rm ee} &= \beta_{\rm ae} \boldsymbol{w}^{\sf H} \boldsymbol{J}_{\rm A} \boldsymbol{w} +  \sigma_{\rm  e } ^ { 2 },\\
            \mathcal{R}_{\rm ae} &= \sqrt{P_{\rm B}}\sqrt{\beta_{\rm ab}}\sqrt{\beta_{\rm ae}}\boldsymbol{w}^{\sf H} \boldsymbol{J}_{\rm A} \boldsymbol{w}=\mathcal{R}^{\sf H}_{\rm ea},\\    
            \mathcal{R}_{\rm be} &= \sqrt{\beta_{\rm ab}}\sqrt{\beta_{\rm ae}}\boldsymbol{w}^{\sf H} \boldsymbol{J}_{\rm A} \boldsymbol{w}=\mathcal{R}^{\sf H}_{\rm eb}.
		\end{align}}
		Then, we can calculate the two determinants in (\ref{eq:MI}) and obtain the KGR as (\ref{eq:whole}).
		%--------------
	  \subsection{Proof of Lemma~2}\label{sec:lemma2}
		First, we denote $\boldsymbol{w}=\sqrt{P}\boldsymbol{w}_0$ with $\|\boldsymbol{w}_0\|^2_2=1$. Since $\frac{\textit{d}R_{\rm SK}}{\textit{d}P} > 0$, the optimal $P$ is {achieved when} $P=P_{\rm A}$. Then, consider the function
        \begin{align}
            f(x)=\frac{\left(P_{\mathrm{B}} x+P_{\mathrm{A}} \sigma^{2}\right)\left(x+\sigma^{2}\right)}{\left(P_{\mathrm{A}} \sigma^{2}+P_{\mathrm{B}} \sigma^{2}\right) x+P_{\mathrm{A}} \sigma^{4}},
        \end{align}
        that is monotonically increasing for $x > 0$, since
        \begin{align}
            \frac{d f(x)}{d x}=\frac{P_{\mathrm{B}}}{\sigma^{2}} \frac{\left(P_{\mathrm{A}}+P_{\mathrm{B}}\right) x^{2}+2 P_{\mathrm{A}} \sigma^{2} x}{\left(\left(P_{\mathrm{A}}+P_{\mathrm{B}}\right) x+P_{\mathrm{A}} \sigma^{2}\right)^{2}}>0 .
        \end{align}
		Denote $x=\beta_{\rm ba}\boldsymbol{w}^{\sf H} \boldsymbol{J}_{\rm A} \boldsymbol{w}$ and the objective function is $R_{\rm SK}=\log_2f(x)$. This completes the proof.

        \subsection{Proof of Lemma~3}\label{sec:lemma3}
        For two complex numbers, $z_1=a_1+ib_1$ and $z_2=a_2+ib_2$, we have
        \begin{align}
            \mathrm{Re}\left(z_1z_2^*\right)&=a_1a_2+b_1b_2\notag\\
            &\leq \sqrt{a_1^2+b_1^2}\cdot\sqrt{a_2^2+b_2^2}.
        \end{align}
        Letting $z_1=f$ and $z_2=f_{l-1}$ yields the desired inequality.
%\begin{comment}
    {
    \subsection{Proof of Lemma~4}\label{sec:lemma4}
    Denote $\boldsymbol{w}=\sqrt{P}\boldsymbol{w}_0$ with $\|\boldsymbol{w}_0\|^2_2=1$. Since $\frac{\textit{d}R_{\rm SK}}{\textit{d}P}$, the optimal $P$ is $P_A$. Then, consider the function

    \begin{align}\label{fx}
            f(P_{\rm A})=\frac{\left(P_{\mathrm{B}} x_0+ \sigma^{2}\right)\left(P_{\rm A}x_0+\sigma^{2}\right)}{\left(P_{\mathrm{A}} \sigma^{2}+P_{\mathrm{B}} \sigma^{2}\right) x_0+ \sigma^{4}},
        \end{align}
    where $x_0=\beta_{\rm ba}\boldsymbol{w}_0^{\sf H}\boldsymbol{J}_{\rm A}\boldsymbol{w}_0$.
        
    To analyze the effect of the maximum power budget $P_{\rm A}$, the derivation of \eqref{fx} with respect to $P_{\rm A}$ yields
      \begin{align}\label{fx_pa_deriavation}
            \frac{\textit{d}f(P_{\rm A})}{\textit{d}P_{\rm A}}
            &=
            \frac{ x_0\left(P_{\mathrm{B}} x_0 + \sigma^2\right) \left(P_{\rm A} \sigma^2 x_0 + P_{\rm B} \sigma^2 x_0 + \sigma^4\right)}{\left(P_{\rm A} \sigma^2 x_0 + P_{\rm B} \sigma^2 x_0 + \sigma^4\right)^2}\notag\\
            &\qquad-\frac{\sigma^2 x_0\left(P_{\mathrm{B}} x_0 + \sigma^2\right)\left(P_{\rm A} x_0 + \sigma^2\right)}{\left(P_{\rm A} \sigma^2 x_0 + P_{\rm B} \sigma^2 x_0 + \sigma^4\right)^2}\notag\\
            &=
            \frac{\left(P_{\mathrm{B}} x_0 + \sigma^2\right) P_{\rm B} \sigma^2 x_0^2}{\sigma^4 \left(P_{\rm A} x_0 + P_{\rm B} x_0 + \sigma^2\right)^2}\notag \\
            &= 
            \frac{\left(P_{\mathrm{B}} x_0 + \sigma^2\right) P_{\rm B} x_0^2}{\sigma^2 \left(P_{\rm A} x_0 + P_{\rm B} x_0 + \sigma^2\right)^2}.
        \end{align}
        
    As $P_{\rm A}$ gradually increases, the denominator of \eqref{fx_pa_deriavation} gradually increases, eventually {forcing} $\frac{\textit{d}f(P_{\rm A})}{\textit{d}P_{\rm A}}$ approaches zero {such that} the value of $R_{\rm SK}$ becomes saturated. {This completes the proof.}
 }
%\end{comment}

\end{appendices}

	\bibliographystyle{IEEEtran}
	\bibliography{IEEEabrv,Ref1}

\begin{IEEEbiography}[{\includegraphics[width=1in,height=1.25in,clip,keepaspectratio]{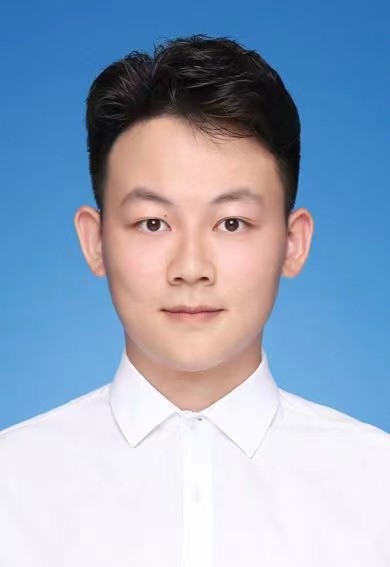}}]{Zhiyu Huang} received the B.S. degree in Communications Engineering and M.S. degree in Signal and Information Processing from Shanghai University, Shanghai, China, in 2021 and 2024. He is currently pursuing the Ph.D. degree with Southeast University. His current research interests include physical-layer security and wireless communication network optimization.
\end{IEEEbiography}

 \begin{IEEEbiography}[{\includegraphics[width=1in,height=1.25in,clip,keepaspectratio]{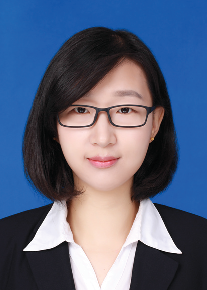}}]{Guyue Li}(S'15-M'17)
received the B.S. degree in Information Science and Technology and the Ph.D. degree in Information Security from Southeast University, Nanjing, China, in 2011 and 2017, respectively. 
From June 2014 to August 2014, she was a Visiting Student with the Department of Electrical Engineering,  
Tampere University of Technology, Finland. 

She is an Associate Professor with the School of Cyber Science and Engineering, Southeast University and Visiting Scholar at Tampere University of Technology, Finland and Université Gustave Eiffel Noisy-le-Grand, France (ESIEE PARIS). Her research interests include wireless network attacks and physical-layer security solutions for 5G and 6G. Her main research topics include secret key generation, radio frequency fingerprint and reconfigurable intelligent surface. She was a recipient of the Young Scientist awarded by International Union of Radio Science (URSI) and won the Youth Science and Technology Prize of Jiangsu Cyber Security Association, the A-Level Zhishan Scholar of Southeast University. Dr. Li has been the Workshop Co-Chair of IEEE VTC from 2021 to 2022. She is currently serving as an Editor of IEEE Communication Letters and an Associate Editor of EURASIP Journal on Wireless Communications and Networking.
\end{IEEEbiography}

\begin{IEEEbiography}[{\includegraphics[width=1in,height=1.25in,clip,keepaspectratio]{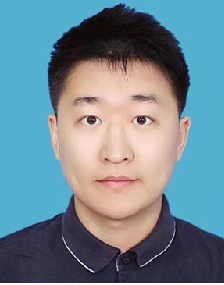}}]{Hao Xu}(S'15-M'19) received the B.S. degree in communication engineering from Nanjing University of Science and Technology, Nanjing, China, in 2013, and the Ph.D. degree in information and communication engineering with the National Mobile Communications Research Laboratory, Southeast University, Nanjing, China, in 2019. From 2019 to 2021, he worked as an Alexander von Humboldt (AvH) Post-Doctoral Research Fellow at the Faculty of Electrical Engineering and Computer Science, Technical University of Berlin, Germany. From 2021 to 2025, he worked as a Marie Sk\l{}odowska-Curie Actions (MSCA) Individual Fellow at the Department of Electronic and Electrical Engineering, University College London, UK. He is currently a professor with the National Mobile Communications Research Laboratory, Southeast University, Nanjing, China. His research interests mainly include communication theory, information theory, mathematical optimization, MIMO systems, and privacy and security. He has been serving as an Associate Editor for IEEE Transactions on Communications since August 2024 and IET Communications since August 2021. He was the recipient of the 2024 IEEE ISTT Best Paper Award.
\end{IEEEbiography}

\begin{IEEEbiography}[{\includegraphics[width=0.95in,height=1.4in]{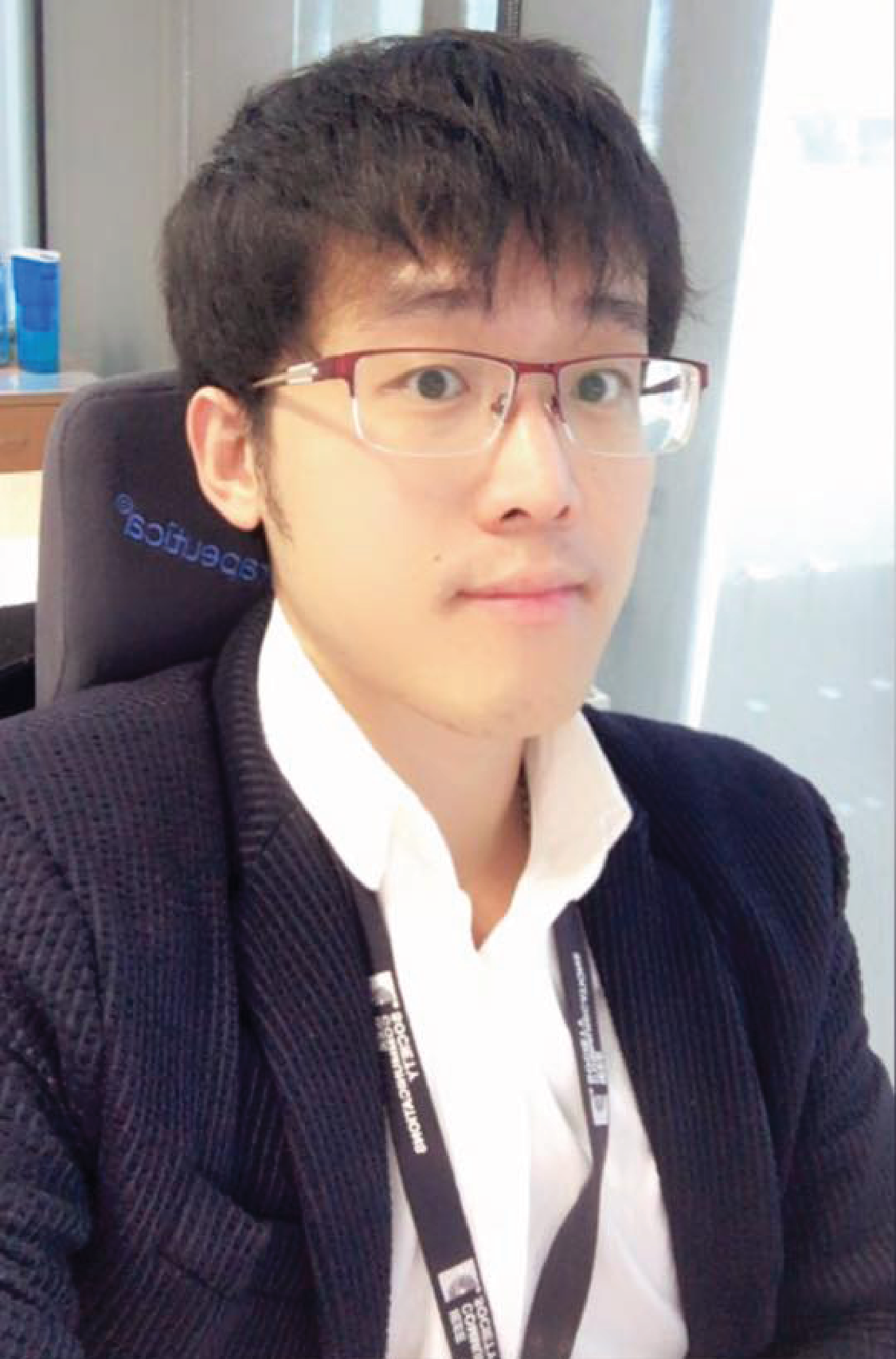}}]{Derrick
Wing Kwan Ng }(S'06-M'12-SM'17-F'21) received a bachelor's degree with first-class honors and a Master of Philosophy (M.Phil.) degree in electronic engineering from the Hong Kong University of Science and Technology (HKUST) in 2006 and 2008, respectively. He received his Ph.D. degree from the University of British Columbia (UBC) in Nov. 2012. He was a senior postdoctoral fellow at the Institute for Digital Communications, Friedrich-Alexander-University Erlangen-N\"urnberg (FAU), Germany. He is now working as a Scientia Associate Professor at the University of New South Wales, Sydney, Australia.  His research interests include global optimization, physical-layer security, IRS-assisted communication, UAV-assisted communication, wireless information and power transfer, and green (energy-efficient) wireless communications. 

Dr. Ng has been listed as a Highly Cited Researcher by Clarivate Analytics (Web of Science) since 2018.  He received the Australian Research Council (ARC) Discovery Early Career Researcher Award 2017, the IEEE Communications Society Leonard G. Abraham Prize 2023,  the IEEE Communications Society Stephen O. Rice Prize 2022,  the Best Paper Awards at the WCSP 2020,  2021,  IEEE TCGCC Best Journal Paper Award 2018, INISCOM 2018, IEEE International Conference on Communications (ICC) 2018, 2021, 2023  IEEE International Conference on Computing, Networking and Communications (ICNC) 2016,  IEEE Wireless Communications and Networking Conference (WCNC) 2012, the IEEE Global Telecommunication Conference (Globecom) 2011, 2021 and the IEEE Third International Conference on Communications and Networking in China 2008.  He served as an editorial assistant to the Editor-in-Chief of the IEEE Transactions on Communications from Jan. 2012 to Dec. 2019. He is now serving as an editor for the IEEE Transactions on Communications and an Associate Editor-in-Chief for the IEEE Open Journal of the Communications Society. 
\end{IEEEbiography}

%\begin{IEEEbiography}[{\includegraphics[width=1in,height=1.25in,clip,keepaspectratio]{fig/Bio/AqHu.eps}}]{Aiqun Hu}
%received the B.Sc.(Eng.), the M.Eng.Sc. and Ph.D. degrees from Southeast University in 1987, 1990, and 1993 respectively. He was invited as a post-doc research fellow in The University of Hong Kong from 1997 to 1998, and TCT fellow in Nanyang Technological University in 2006. 

%He is currently a Professor at the National Mobile Communications Research Laboratory, Southeast University. 
%His research interests include data transmission and secure communication technology. He has published two books and over 100 technical papers in wireless communications field.
%\end{IEEEbiography}

\end{document}